\documentclass[referee]{aa}
\usepackage{graphicx}
\usepackage{longtable}
\usepackage{natbib}          
\bibpunct{(}{)}{;}{a}{}{,}
\usepackage{txfonts}
 \defcitealias{MB08}{Paper~I}
\begin{document}

\title{Dust properties along anomalous extinction sightlines.}
\subtitle{II. Studying  extinction curves with dust models}

\author{P. Mazzei\inst{\ref{inst1}} \and  G. Barbaro\inst{\ref{inst2}}  }

\institute{ INAF,  Astronomical Observatory, Vicolo dell'Osservatorio, 5
    Padova, 35122, Italy\\
	\email{paola.mazzei@oapd.inaf.it}\label{inst1}
\and
Department of Astronomy, Vicolo dell'Osservatorio 3, 35122 Padova, Italy \label{inst2}
}
\date{Received ; accepted}

\abstract 
{Recent works devoted to study the extinction in our own Galaxy pointed out that the large majority of sight lines analyzed obey  a simple relation depending on one parameter, the total-to-selective extinction 
coefficient, $R_V$. Different values of $R_V$ are able to match the whole  
extinction curve through different environments so characterizing the normal extinction curves.
However, as outlined in several recent papers, anomalous curves i.e, curves which strongly deviate from such simple behavior do exist in our own Galaxy as well as in external galaxies.}
{In this paper more than sixty  curves with large ultraviolet deviations from their best-fit one parameter curve are analyzed. 
The extinction curves are fitted with dust models to shed light into the properties of the grains along selected lines of sight,  the processes  affecting them, and  their relations with the environmental
characteristics. }
{The extinction curve models are reckoned by following recent
prescriptions on grain size distributions able to 
describe one parameter curves for R$_V$ values  from 3.1 to 5.5. Such models, here extended down to R$_V$=2.0,  allow us to compare the resulting properties of our deviating curves with the same as normal curves in a self-consistent framework, and thus to recover the relative trends overcoming the modeling uncertainties.
}
{Together with twenty anomalous curves extracted from the same sample, studied in a previous paper  and here revised to account for recent updating,
such curves represent the larger  and homogeneous sample of anomalous curves studied so far with  dust models. 
Results show that the ultraviolet deviations are driven  by a larger amount of small grains than predicted for lines of sight where extinction depends on one parameter only. Moreover, the dust-to-gas ratios  of anomalous curves are lower than the  same values for no deviating lines of sight.}
{Shocks and grain-grain collisions should both destroy dust grains, so reducing the amount of the dust trapped into the grains,  and modify
the  size distribution of the dust, so increasing the small-to-large grain size ratio. Therefore, the  extinction properties derived should arise along sight lines where shocks and high velocity flows perturb the physical state of the interstellar medium living their signature on the dust properties.}

	
\titlerunning{Anomalous extinction curves}
\authorrunning{P. Mazzei \& G. Barbaro}

\maketitle

\section{Introduction}

The picture of the extinction in the Galaxy is very complex, with large variation from region to region. A successful attempt to interpret the observed behavior of extinction curves in our own Galaxy from near-IR to ultraviolet (UV)  was given by  \citet[CCM in the following]{CCM89}. They found   a relation between the whole shape of the extinction curve and the total-to-selective extinction 
coefficient, $R_V$. With only very few exceptions, Galactic extinction curves observed so far tend to follow this relation within the uncertainties of the calculated $R_V$ values and extinction curves (\citet{C00}, \citet{Goetal03}, \citet{Vaetal04} and references therein). Different values of R$_V$ are a rough indicator of different environmental conditions which affect the grain size distribution: low-R$_V$ values arise along sight lines with more small grains than high-R$_V$ sight lines.
However, as pointed out   by \citet{CC91}, \citet{Maca92}, and in several recent papers (\citet{MB08}, and references therein) anomalous curves, i.e., curves which deviate from this simple behavior, still exist in our own Galaxy. \citet{Vaetal04} found that seven per cent of their sample of 417 {\it International Ultraviolet Explorer} (IUE) extinction curves combined with Two-Micron All-Sky Survey (2MASS) photometry, deviate from the CCM law by more than three times the standard deviation (3$\sigma$).
Moreover the CCM law does not apply outside the Galaxy. \citet{Goetal03} showed that the large majority of measured extinction curves in the Large and Small Magellanic Clouds do not obey the CCM law, even if  a continuum of dust properties exists. 
\citet{FM09} recently concluded that to fit the visible-infrared region of the extinction curve two parameters are needed, at least. The power-law model for the near-IR extinction law provides an excellent fit to most extinction curves, but the value of the power index varies significantly from sight line-to-sight line and increases with the wavelength.\\
The interest into these problems is rising since  suitable extinction corrections, which allow to properly account for galaxy properties (i.e., colors and luminosities of nearby as well as of distant galaxies) are needed to improve our  knowledge of galaxy evolution. Thus, our understanding of the dust extinction
properties, in particular of their dependence on the environment, are challenges
to modern cosmology.\\
In this paper we deepen the analysis of \citet{MB08} (Paper~I in the following) by studying the behavior of a new class of extinction curves singled out from the same sample  just defined in that paper where 
 785 extinction curves have been compared with the relations derived by CCM  for a variety of R$_V$ values in the range 2-6.
The curves have been classified as normal if they fit at least one of the CCM curves or anomalous otherwise. In particular,
all the curves retained deviate by more than $2\sigma$ 
from their  CCM best-fit law, at least at one UV wavelength.
By fitting the observed data with  extinction curves provided by dust grain models, we aim at giving insight into
the properties of the grains along selected lines of
sight,  the processes  affecting them, and  their relations with the environmental
characteristics. 
The extinction curve models are reckoned by following the 
prescriptions of \citet{WD01} i.e., using their grain size distributions 
together with the more recent updating \citep{DL07}. 
 Models of \citet{WD01},  able to
describe  normal curves for R$_V$ values 3.1, 4.0 and 5.5, have been  extended here down to  R$_V$=2.0 and updated following \citet{DL07}. All such models  allow us to compare the resulting properties, both of normal and of anomalous curves,  in a self-consistent framework, and thus to recover the relative trends.

The plan of the paper is the following: section 2 summarizes the sample of extinction curves
and the method used to derive their anomalous behavior,  more details into these points are given in \citetalias{MB08}; section 3 is devoted to the dust models  built up to best-fit the selected curves.
 All the models,  for both anomalous  and normal curves, are computed using the grain size distributions of
\citet{WD01} with the more recent updating \citep{DL07}; we will indicate such models as  WD in the following. 
In section 4  the results from all such models are presented in terms of dust-to-gas ratios, abundance \footnote{By "abundance", we mean the number of atoms of an element per interstellar H} ratios and small-to-large grain size ratios of the dust trapped into the grains along extinction curves.
Results from Paper 1 are also revised accounting for the previous mentioned implementation, to allow the comparison of the properties of the whole sample of anomalous sightlines.
Section 5 is devoted to 
the \citet{FM88, FM90}
parameterization of  all our  models. The aim is to compare the properties of our sample  with those of
the larger sample of parameterized sightlines in  literature  available so far \citep{Vaetal04}. In section 6 there are  conclusions.

\section{The sample}  

The source of the UV data  is the {\it Astronomical Netherlands Satellite} (ANS) photometry catalog of \citet{Weetal82}.
The UV observations  were performed in five UV bands with central wavelengths (and widths) 1549 (149), 1799 (149), 2200 (200), 2493 (150), and 3294 (101)\,\AA.
Of the approximately 3500 stars in the ANS catalog, \citet{Saetal85} derived UV extinction excesses for 1415 normal stars with spectral type earlier than B7. These  color excesses (Table 1 of \citet{Saetal85}), E$(\lambda-$V)  for $\lambda$ cited above, are referenced to the  photoelectric V band, starting from UV magnitudes listed in the ANS catalog and intrinsic colors by \citet{Wuetal80}; absolute calibrations of UV fluxes were performed as described by \citet{Weetal82}; E(B-V) data are also listed in the Table 1 of the same catalog.

To avoid large errors in the color excesses, only those lines of
sight with E(B-V)$\ge$ 0.2  have been retained, amounting to   785 curves.  From such a sub-sample, \citet{Baetal01} singled  out 78 lines of sight which they defined as anomalous.
Their  analysis were extended in \citetalias{MB08} by considering near-IR magnitudes from 2MASS catalog to derive the 
intrinsic infrared colors indices by using Wegner's calibrations (1994).
For  each observed curve covering the IR and UV region, the following quantities have been minimized through a weighted  least square fit with different  standard CCM relations:

\begin{equation}
  \delta (x_r)_{i,j}=[ \kappa (x_r)_{j} - \kappa (x_r)_{i}] / \sigma_{\kappa(x_r)_i}  
\end{equation}

 where the index $r$ refers to all the eight wavelengths i.e., the five UV 
ones from the \citet{Saetal85}, cited above,  and  the three IR wavelengths;
setting x$_r$=1/$\lambda_r$ and
$\kappa(x_r)=E(\lambda_r-V)/E(B-V)$, $\kappa( x_r)_{i}$
refer to the observed curves (index i) and $\kappa (x_r)_{j}$ to the
CCM curve corresponding to each $R_V$ value ranging from $\simeq$2 to 6.
The $(\sigma_{\kappa(x_r)_i} $, are computed following eq. (3) of  \citet{We02} :
\begin{equation}
\sigma_{\kappa(x_{r})}^2=[\frac{1}{E(B-V)}]^2 [\sigma^2_{m_{\lambda_r}}
 +\sigma_{V}^2+\sigma^2_{r,m}]+[\frac {\kappa(x_{r})\sigma_{E(B-V)}}
 {E(B-V)}]^2
\end{equation}
 accounting for i) a conservative maximum color excess
error, $\sigma E(B-V)$, of 0.04 mag,  ii) a root-mean-square deviation of the
observation at .55\,$\mu$m, $\sigma_{V}$,  of 0.01 mag \citep{Saetal85},
iii) a root-mean-square deviation of the photometric observation at wavelength
$\lambda_r$, $\sigma_{m_{\lambda_r}}$, ranging from 0.001--0.218 mag \citep{Weetal82}
in the UV range and given by the 2MASS catalog in the IR one,
iv) classification errors and errors in the
intrinsic colors i.e, $\sigma_{r,m}$, as given in Table 1B of \citet{MS81} in the UV range and
as derived from \citet{We94} in the IR one.

For each  line of sight,
the residual differences   at the five UV wavelengths
between the observed data and the  best-fit  standard CCM curve, as in eq. (1),  have been evaluated.
Only  those lines for which at least one of such $|\delta_{r}|$ exceeds (or equals) two
have been retained. This  defines the $2\delta$ sample.\\ 
As pointed out in \citetalias{MB08}, such an approach is different from that of \citet{Baetal01},  both because  IR data were not yet available, and because the anomalous character is defined here by
analyzing separately  each UV  wavelength,  i.e., considering
$\delta_r$, instead  of   using  a criterion 
based on the combination of all the UV data (i.e., $\Delta^2$, see  \citet{Baetal01} for more 
details). 
There are 84   lines of sight in such a new sample  i.e., more than 10\% of the
selected initial sample, 27 (3.4\%) with at least one $|\delta_{r}| > $3. 
Such percentages are higher than those expected from random error analysis,  as pointed out in \citetalias{MB08}.

The five UV wavelengths cited above correspond to the
values: x=6.46, 5.56, 4.55, 4.01, and 3.04, respectively.
The behavior of $\delta(4.55)>$0, more or less in correspondence of the bump, and of  $\delta(6.46)<$0,  in correspondence of the 
far-UV rise,  was shown  for the whole sample in Fig. 1 of \citetalias{MB08}.
There are fifteen sight lines with $\delta(4.55)>$0 and  $\delta(6.46)<$0, called type A anomalous  curves,
analyzed in \citetalias{MB08} together with five sight lines with $\delta(4.55)<$0 and  $\delta(6.46)>$0,
defined  type B anomalous  curves.
Type A curves are characterized by weaker bumps and steeper  far-UV rises  than
expected from their best-fit CCM curve,  worse  by more than 2$\sigma$ at least  one UV wavelength for each of them.
Type B  curves show stronger bumps  together with smoother far-UV rises than
expected for CCM curves which best-fit observations.

\begin{figure*}
\centering
\includegraphics[width=6.0cm]{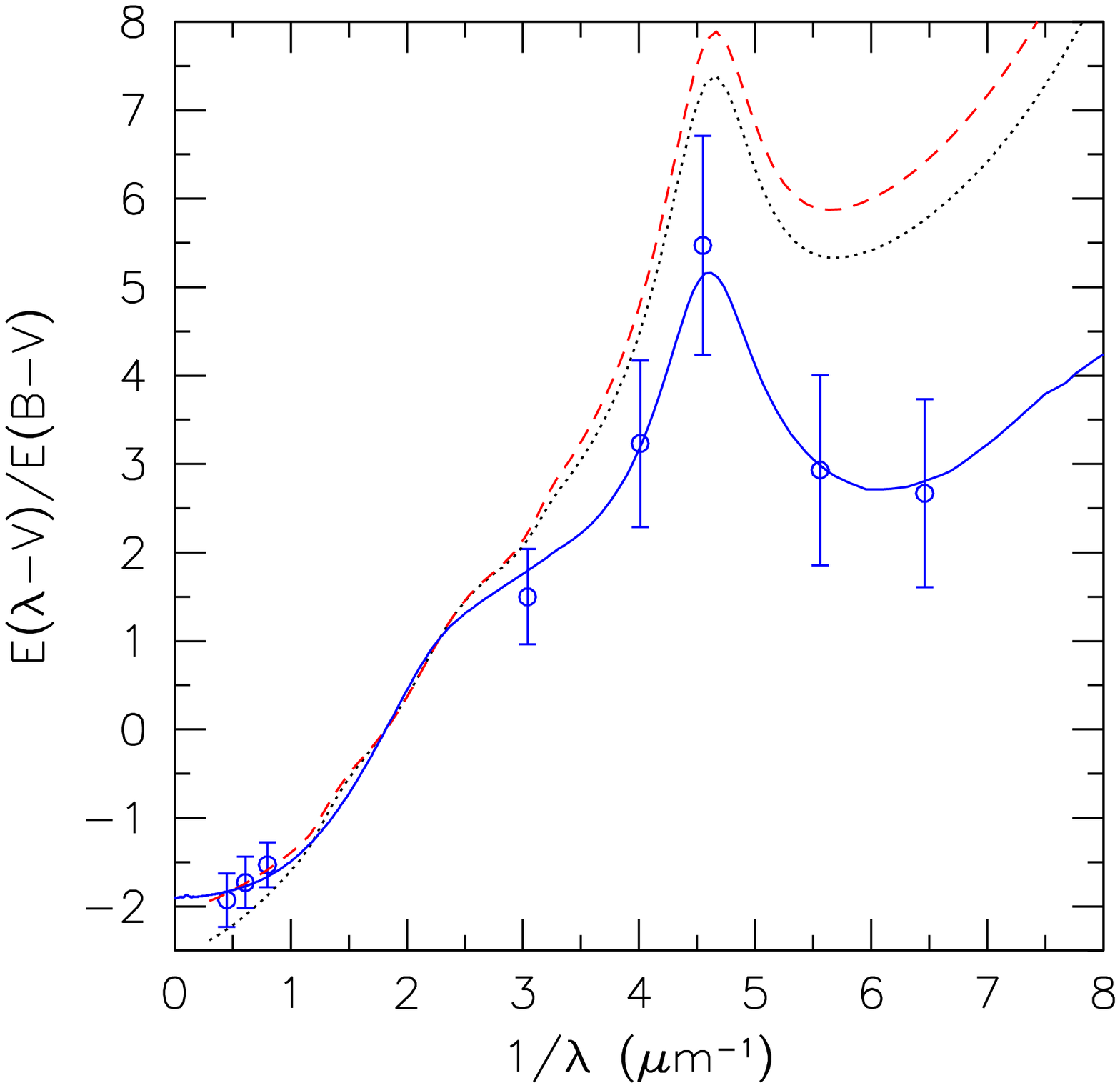}
\includegraphics[width=6.0cm]{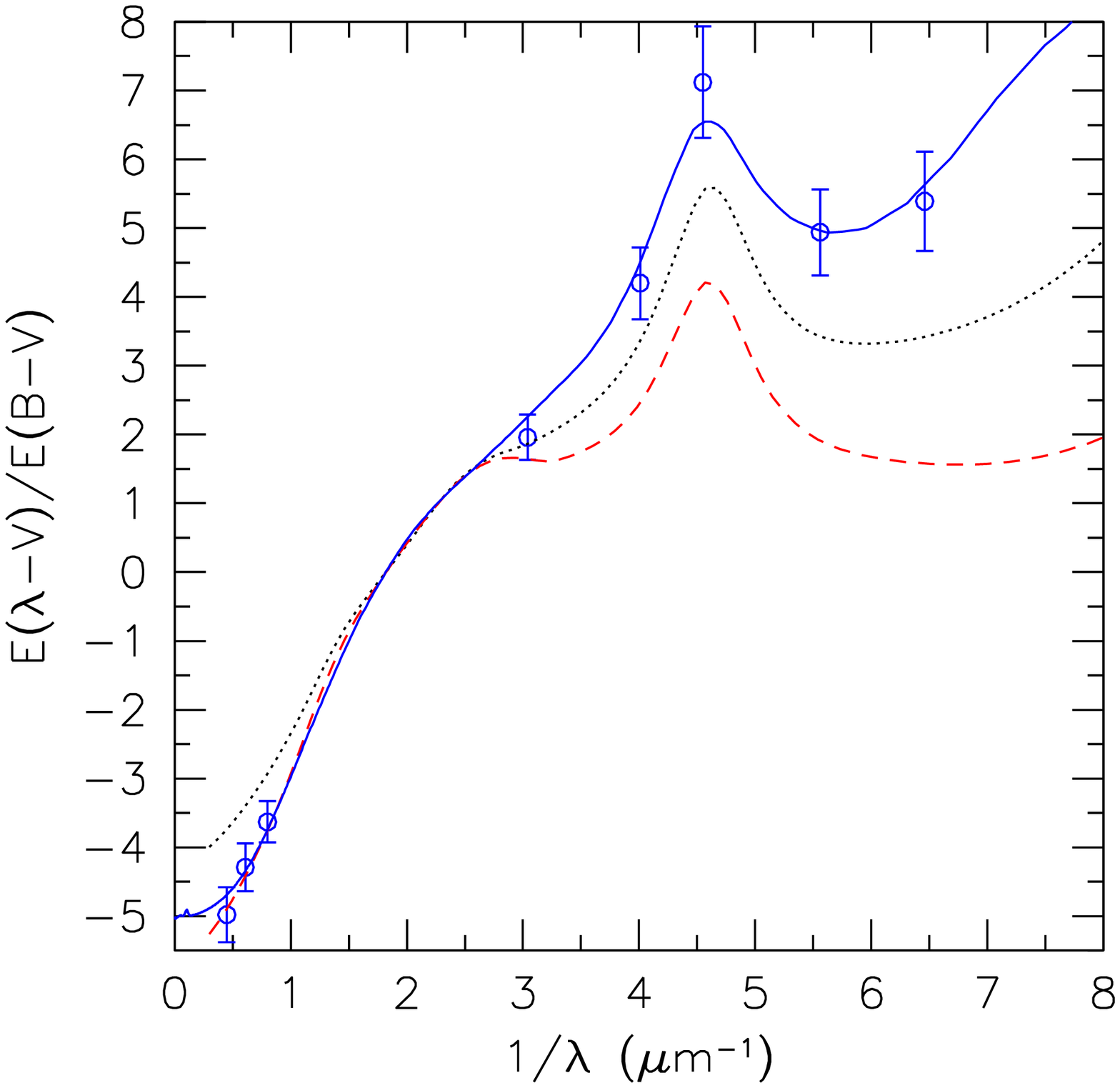}

 \caption{
 {\sl Left panel:} Extinction data of  HD~168785 (open circles) compared with our best-fit   dust model (see Section 3),(blue) continuous line, and the best-fit  standard CCM curves corresponding to the whole spectral range (black) dotted line, and to the IR data only, (red) dashed line.
 {\sl Right panel:} Extinction data of   BD+58\,310 (open circles) compared with our  best-fit   dust model (blue) continuous line, and the corresponding best-fit   standard CCM curves with the same symbols as in the left panel.
}
\label{anomB_sav}
\end{figure*}

Here we focus on  type C  curves i.e., sixty-four lines of sight, 76\% of $2\delta$  sample. 
The behavior of type  C curves can be also recovered here by looking at  
Fig. \ref{anomB_sav}. 
For all such curves, with the exception of five curves only,  the corresponding best-fit  standard  CCM curve is always well above ($\ge$ 2$\sigma$ at least one UV wavelength) the observed  data (Fig. \ref{anomB_sav}, left-panel).
For five of them, two belonging to the 2$\delta$ sample, i.e., BD+59\,2829 and BD+58\,310, and three  to the 
3$\delta$ sample, i.e., HD\,14707,   HD\,282622, and  BD+52\,3122,  the corresponding best-fit  standard CCM curve 
is well below the observed data, with some exception at x=3.04 (Fig. \ref{anomB_sav}, right-panel).

In Table \ref{anom0}
the main properties of type C  curves  are presented: names
(col. I), spectral types (col. II), reddening (col. III), V magnitudes (col IV), all  from \citet{Saetal85}, and
R$_V$ (col. V).
The R$_V$ values in Table 1, obtained  by minimizing through the weighted  least square fit with different CCM relations the quantities in eq. (1) using IR the extinction data only,  agree 
with estimates of  R$_V$ following  prescriptions of \citet{F99}; the errors  have been computed with the same method
used in \citetalias{MB08}  \citep{GP04},
which accounts for mismatch errors affecting the color excesses to the larger extent.\\
The first twenty two lines of sight in Table 1 belong to the 3$\delta$ sample.\\
We notice that  HD\,392525 corresponds to BD$+$57\,2525 and  HD\,282622  to BD$+30$\,748.

From {\it SIMBAD} CDS database (http://simbad.u-strasbg.fr/simbad)  about 8$\%$ of our sight lines correspond to Be stars (i.e., HD\,21455, HD\,28262, HD\,326327, HD\,392525, and BD$+$59\,2829), and 9$\%$ to variable  stars (i.e., HD\,1337, HD\,14707, HD\,28446, HD\,141318, HD\,217035, BD$+$31\,3235); these represent about 17$\%$ of type C curves.

In the following analysis we removed from the sample BD$+$56\,586, since  a negative color excess: E(.33-V), -0.01,  corresponds  to such line of sight, that suggests  mismatch errors (\citet{Paetal01},\citet{We02}).
We also removed from the sample 
HD\,1337, an eclipsing binary of $\beta$ Lyrae type in {\it SIMBAD}, marked as a variable in \citet{Saetal85} too, and HD\,137569,  a post-AGB star ({\it SIMBAD}). Their extreme R$_V$ values,  0.60$\pm$0.18 and 1.10$\pm$0.18 (Table 1) are outside the range of explored  standard CCM relations \citep{F99}.

The selected lines of sight probe a wide range of dust environments, as suggested by the  
large spread of the measured  $R_V$ values. 
In particular, HD\,141318 and BD+58\,310  are  extreme cases:
HD\,141318 with R$_V$ equal to 1.95$\pm$0.18 and BD+58\,310 
with R$_V$=5.65$\pm0.47$ (Table \ref{anom0}). 

\addtocounter{table}{1}


\section{Models}

Models of the extinction curves have been computed according to the prescriptions
of \citet{WD01} and \citet{DL07}.
The novelty of such models is that the  grain size distribution is
 based on  more recent observational constraints in the optical and infrared spectral domains
 (see \citet{WD01} and references therein).
 Such  distribution,  a revision of the  
 \citet{MRN}  size distribution (see \citet{Cletal03}  for a discussion on grain
 size distribution history),
accounts for two populations of spherical grains: amorphous
silicate (Si) and carbonaceous grains (C), the latter consisting of graphite grains 
 and polycyclic aromatic hydrogenated (PAH) molecules.
Their optical properties, which depend on their
geometry and chemical composition, are described by  \citet{lid01a, lid01b} and
by \citet{DL07}, and account for new laboratory data i.e., the near-IR absorption spectra measured by \citet{Matetal05},  as well as the spectroscopic  observations of PAH emission from dust in nearby galaxies 
(\citet{DL07}, and references therein).

\begin{figure*}
\centering
\includegraphics[width=5.50cm]{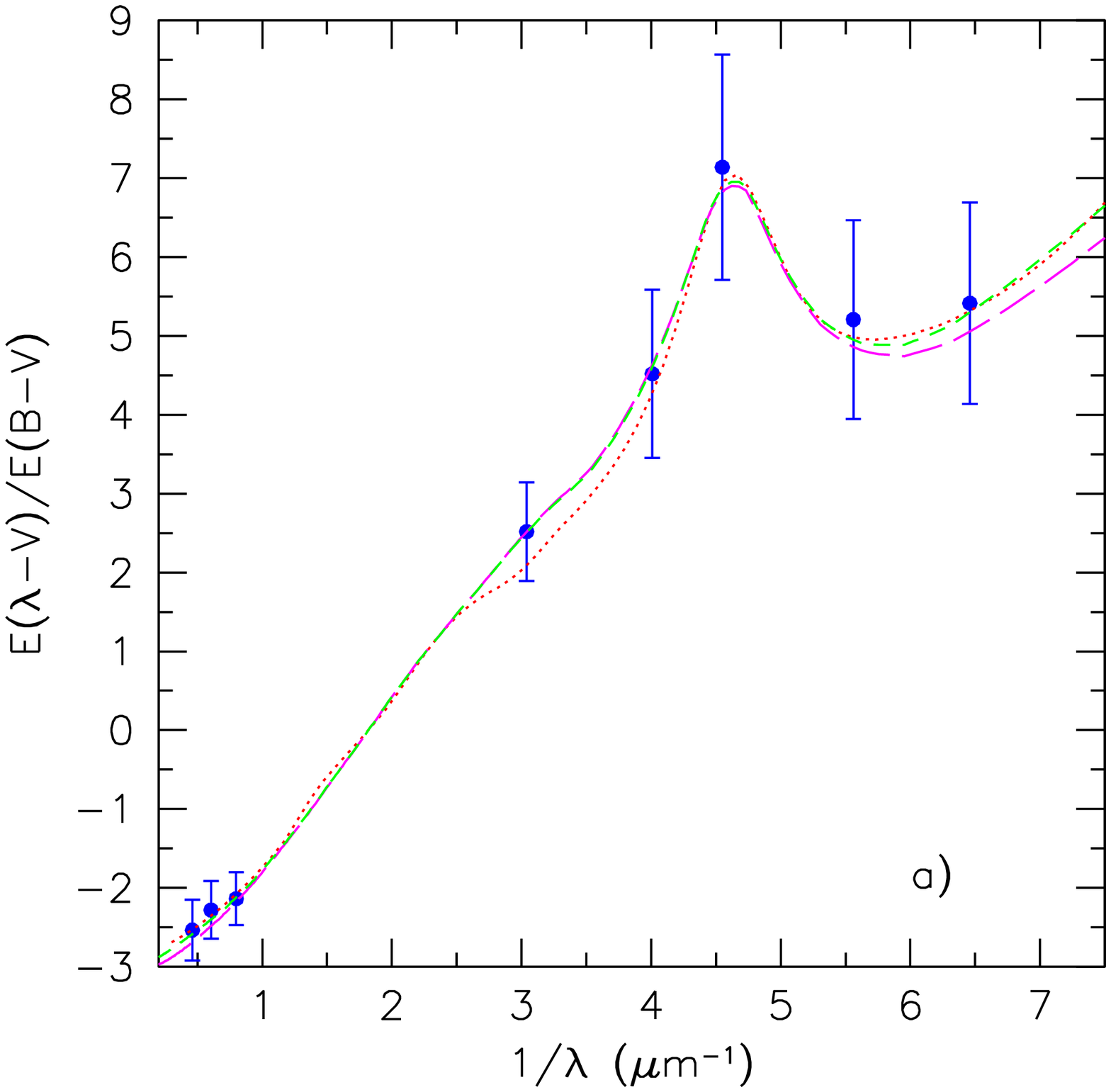}
\includegraphics[width=5.50cm]{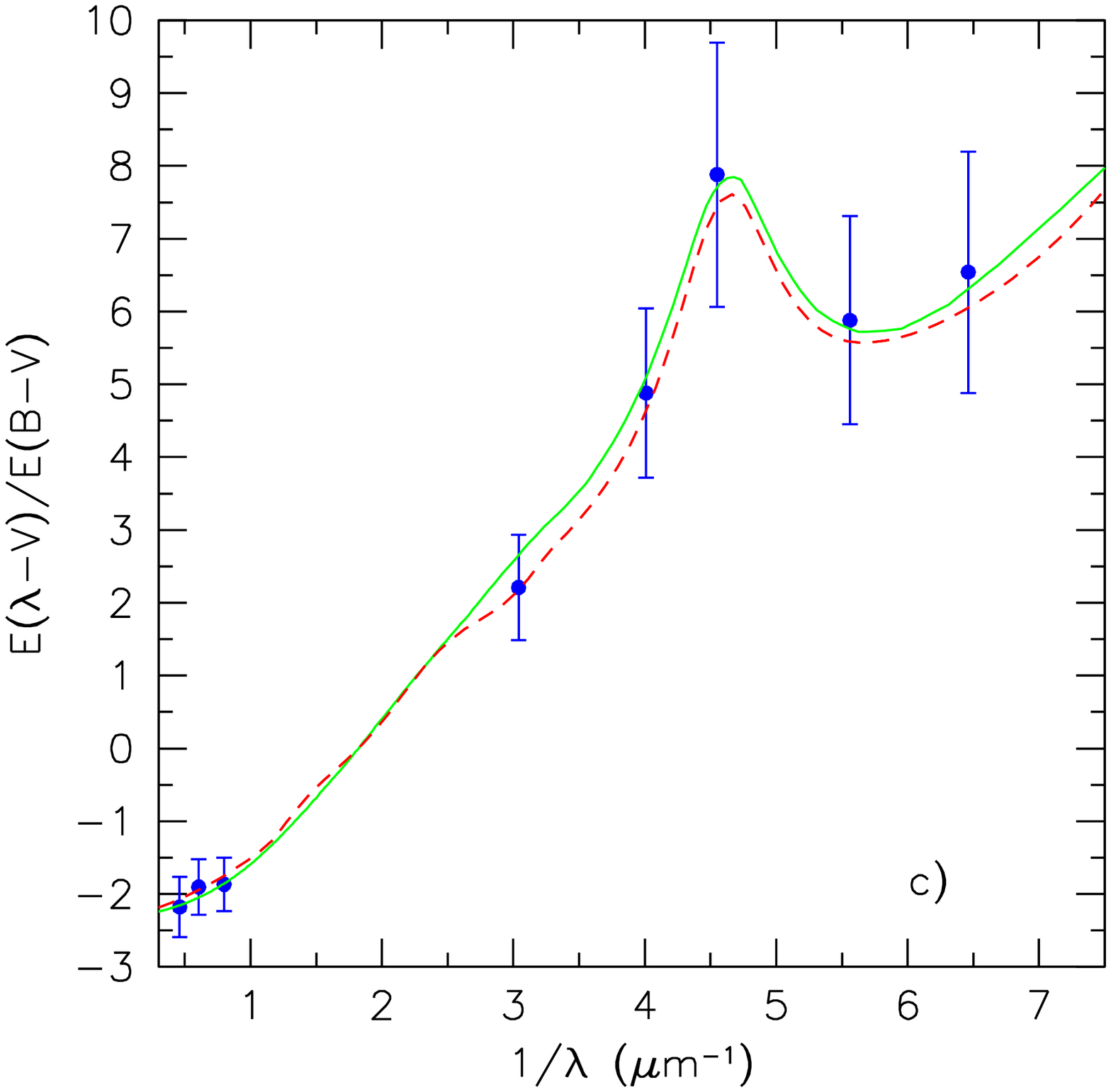}
 \caption { 
 {\sl Panel a):} HD\,39136,  a normal curve of our sample (blue filled squares) whose  best-fit standard CCM curve corresponds to R$_V$= 2.85$\pm$0.47 (red dotted line). The (magenta) long-dashed line is computed as the seventh  model in Table 1 of \citet{WD01}, R$_V$=3.1,  including latest updating \citep{DL07}; the (green) short-dashed  line is our best-fit dust model: the  fifth model in Table \ref{newWD01}.
{\sl Panel b):} HD\,12323, a normal curve of our sample (blue filled squares) whose  best-fit standard CCM
curve corresponds to R$_V$= 2.30$\pm$0.51 (red dashed line); the (green) continuous line is the best-fit dust model:  the thirteenth  model in Table \ref{newWD01}.
}
\label{newfig}
\end{figure*}

The \citet{WD01} size distribution (see their eq. 4) 
allows both for a smooth cutoff for grain size $a>a_t$, 
and for a change in the slope $d\, ln\, n_{gr}/d\, ln\, a$ for $a<a_t$.
This requires several parameters which can be determined by the comparison with  the observed
curve, the comparison being performed with the Levenberg-Marquardt algorithm \citep{WD01}. 
\footnote {In order to allow the method to work, the number of points have been increased so that each curve comprises  hundreds points: nine points equally spaced in $\lambda^{-1}$  are added in each wavelength range  down to zero.}
Table \ref{parmod} 
presents the  size distribution parameters derived from our best-fit  dust models of type C extinction curves where: b$_C$ is the 
abundance of carbon (per H nucleus) in the double log-normal  very small grain population  (see Table 2 and eq. 12-14 of \citet{DL07}), 
$\alpha_g$ and  $\alpha_s$ are the power law indexes of carbon and Si
grain size distributions respectively, $\beta_g$ and $\beta_s$ their
curvature parameters, a$_{t,g}$ and a$_{t,s}$ their transition sizes, a$_{c,g}$ and a$_{c,s}$ their upper cutoff radii.
\addtocounter{table}{1}
Such models, indeed, depend on ten parameters since a$_{c,s}$  results to be 
constant (\citet{WD01}, \citetalias{MB08}). 
Fig. \ref{anomB_sav} compares the observed extinction curves of two type  C lines with the corresponding best-fit models.\\
\indent The values of such parameters able to reproduce the observed wavelength-dependent extinction law in the local Milky Way (MW),   i.e. the observational fits of \citet{F99} for R$_V$ values 3.1, 4.0, and 5.5  and different b$_C$ amounts, corresponding to
twenty-five models,  are derived by \citet{WD01}  (Fig.s 8-12; see also \citet{D03}). 
\citet{LD02} showed that these grain models are also consistent with the observed IR emission from diffuse clouds in the MW and in the SMC.
 The observational fits of \citet{F99}, in their turn, well agree with standard CCM curves for the same R$_V$ values until R$_V$ is smaller than 5.5 \citep[his Fig. 7]{F99}. 
Moreover, as pointed out in Paper 1,  the  extinction curves of \citet{WD01} are almost unaffected by taking into account  the more recent updating \citep{DL07}.
Thus, WD models are useful tools to give insight into the properties of the dust trapped into the grain along normal  lines or small deviating extinction lines (i.e. $<$2 $\sigma$).
 However, since the majority of  R$_V$ values in Table \ref{anom0} are smaller than 3.1,  a new set of WD parameters able to fit normal, CCM, curves with total-to-selective extinction coefficients from 2.9 and 2.0, have been computed and listed in Table \ref{newWD01} for each pair of values (R$_V$, b$_C$), as in   \citet{WD01}.
The last two columns of such a Table  show volumes of carbonaceous and silicate populations normalized to their abundance/depletion-limited values, i.e. 2.07$\times$ 10$^{-27}$\,cm$^{-3}$\,H$^{-1}$ and   2.98$\times$ 10$^{-27}$\,cm$^{-3}$\,H$^{-1}$, respectively \citep{WD01}.
As discussed by \citet{D03} and \citet{D04},
models in Table 1 of \citet{WD01}  slightly exceed the  abundance/depletion-limited values of
 silicon grains ($\le$20\%) and we assume such a value as the maximum allowed. Fig. \ref{newfig} shows two normal curves of our sample with small R$_V$, their best-fit standard CCM curves and, as a comparison, WD dust grain models to fit the data.
Properties of all these models, both of normal and of  anomalous curves,  are given in the next section in terms of dust-to-gas ratios, abundance ratios, and small-to-large grain size ratios of the dust trapped into the grains along such extinction curves.\\
\indent There are several distinct interstellar dust models that simultaneously fits the observed extinction, infrared emission, and abundances constraints \citep{Zuetal04}.
Models of WD 
allow us to compare the results of our deviating curves here with those of normal curves, in a self-consistent framework.

\begin{table*}
\caption{Best-fit parameters of WD grain size distributions for CCM curves with small R$_V$ values}
\label{newWD01}
\centering
\begin{tabular}{lllllllllllll}
\hline\hline
R$_V$&$10^5b_C $ & $\alpha_g$&$\beta_g$&a$_{t,g}$&a$_{c,g}$&C$_g$&$\alpha_s$&$\beta_s$&a$_{t,s}$ &C$_s$&V$_g^C$&V$_g^{Si}$\\
   &      &       &     &$(\mu m)$&$(\mu m)$&  &  &           &$(\mu m)$&&&\\
\hline

2.9 &0.0&-1.90&-0.97&0.008&0.450&1.96$\times 10^{-10}$ &-2.32&0.35&0.168&7.85$\times 10^{-14}$ &0.74& 1.04\\
2.9 &1.0&-1.96&-0.71&0.007&0.650&1.62$\times 10^{-10}$ &-2.32&0.47&0.166&7.50$\times 10^{-14}$ &0.66& 1.04\\
2.9 &2.0&-1.90&-0.51&0.006&0.597&1.82$\times 10^{-10}$ &-2.32&0.47&0.168&7.85$\times 10^{-14}$ &0.80& 1.11\\
2.9 &3.0&-1.77&-0.20&0.008&0.760&2.70$\times 10^{-11}$ &-2.36&0.60&0.153&7.50$\times 10^{-14}$ &0.82& 0.98\\
2.9 &4.0&-1.75&-0.23&0.008&0.760&4.00$\times 10^{-11}$ &-2.37&0.60&0.163&7.50$\times 10^{-14}$&1.00&1.17\\
2.6 &0.0&-2.45&-0.12&0.013&0.705&2.10$\times 10^{-11}$ &-2.41&0.39&0.148&7.00$\times 10^{-14}$ &0.55& 0.79\\
2.6 &1.0&-2.36&-0.10&0.012&0.500&2.10$\times 10^{-11}$ &-2.41&0.39&0.155&7.00$\times 10^{-14}$ &0.60& 0.90\\
2.6 &2.0&-2.30&-0.08&0.012&0.500&1.51$\times 10^{-11}$ &-2.40&0.39&0.155&7.00$\times 10^{-14}$ &0.56& 0.89\\
2.6 &3.0&-2.30&-0.04&0.012&0.250&1.51$\times 10^{-11}$ &-2.40&0.44&0.165&7.00$\times 10^{-14}$ &0.57& 1.07\\
2.6 &4.0&-2.15& 0.01&0.012&0.187&1.01$\times 10^{-11}$ &-2.40&0.44&0.165&7.00$\times 10^{-14}$ &0.57& 1.07\\
2.6 &5.0&-2.10& 0.03&0.012&0.187&1.01$\times 10^{-11}$ &-2.40&0.39&0.158&8.75$\times 10^{-14}$ &0.71& 1.16\\
2.3 &0.0&-2.65&-0.18&0.014&0.705&1.65$\times 10^{-11}$ &-2.40&0.32&0.158&7.00$\times 10^{-14}$ &0.41& 0.90\\
2.3 &0.5&-2.60&-0.15&0.014&1.050&1.50$\times 10^{-11}$ &-2.420& 0.22&0.155&1.50$\times 10^{-13}$ &0.47&0.96\\
2.3 &1.0&-2.75&-0.17&0.012&0.705&1.50$\times 10^{-11}$ &-2.42&0.42&0.155&7.00$\times 10^{-14}$ &0.42& 0.91\\
2.3 &2.0&-2.55&-0.10&0.011&0.705&1.65$\times 10^{-11}$ &-2.30&0.34&0.158&7.00$\times 10^{-14}$ &0.36& 0.90\\
2.3 &3.0&-2.15&-0.27&0.012&0.700&1.60$\times 10^{-11}$ &-2.44&0.48&0.147&8.00$\times 10^{-14}$ &0.48& 0.93\\
2.3 &4.0&-2.10&-0.20&0.010&0.980&1.97$\times 10^{-11}$ &-2.44&0.45&0.145&9.80$\times 10^{-14}$ &0.62& 1.10\\
2.3 &5.0&-2.10&-0.08&0.011&0.988&1.13$\times 10^{-11}$ &-2.47&0.65&0.143&9.80$\times 10^{-14}$ &0.69& 1.17\\
2.0 &0.0&-2.12&-1.86&0.009&0.105&9.30$\times 10^{-11}$ &-2.37&0.09&0.146&5.60$\times 10^{-14}$ &0.17& 0.49\\
2.0 &1.0&-2.00&-0.99&0.010&0.150&2.98$\times 10^{-11}$ &-2.38&0.03&0.147&5.60$\times 10^{-14}$ &0.24& 0.48\\
2.0 &2.0&-1.95&-0.50&0.009&0.450&1.50$\times 10^{-11}$ &-2.46&0.37&0.136&6.00$\times 10^{-14}$ &0.24& 0.55\\
2.0 &3.0&-1.95&-0.30&0.008&0.450&1.30$\times 10^{-11}$ &-2.52&0.65&0.136&6.00$\times 10^{-14}$ &0.27& 0.65\\
2.0 &4.0&-1.95&-0.30&0.010&0.450&1.30$\times 10^{-11}$ &-2.65&1.25&0.136&6.00$\times 10^{-14}$ &0.42& 0.92\\
2.0 &5.0&-1.95&-0.30&0.013&0.450&1.30$\times 10^{-11}$ &-2.70&1.70&0.136&6.00$\times 10^{-14}$ &0.46& 1.09\\

\hline
\end{tabular}
\end{table*}

\section {Results}

For a given  chemical composition of the grains, the total volume occupied by the dust grains, $V_g$,  which includes
the total volume per hydrogen atom of each grain type,  is directly connected with
the dust-to-gas ratio, $\rho_d/\rho_H$, as described in eq.s 3, 4, 5, and 6 of \citetalias{MB08} i.e.:

\begin{equation}
\frac{\rho_{d}}{\rho_{H}}= \frac{\rho_{g}}{m_{H}} \frac{V_{g}}{n_{H}}=
\frac{4\pi}{3m_HN_H} \sum_X \rho_X \int_{a_{min}}^{a_{max}}a^3N_X(a)da
\label{for 222} 
\end{equation}

where the integral is computed for the size distributions of the grains, i.e., from parameters in Table \ref{parmod}
for anomalous curves  and, for normal curves, both in Table 1 of \citet{WD01}, however accounting for latest updating \citep{DL07}, and in Table \ref{newWD01}.

The number of atoms of an element per  interstellar H nucleus
trapped in the grains i.e, the C/H and Si/H  abundances,
and  the same fractions compared with the solar values i.e., C/C$_{\odot}$ and 
Si/Si$_{\odot}$, can be computed from the $\rho_{d}/\rho_{H}$ ratio of each type of grains.
The solar values of carbon and silicon dust abundances 
adopted at this aim are, as in \citetalias{MB08}, (C/H)$_{\odot}=3.3\times 10^{-4}$, 
(Si/H)$_{\odot}=3.65\times 10^{-5}$;  the average mass in our own galaxy of one  
carbon grain is 19.93$\times$10$^{-24}$\,gr/cm$^{-3}$ and that
of one silicon grain   28.7$\times$10$^{-23}$\,gr/cm$^{-3}$, as in \citet{WD01};
such values are in good agreement with  recent estimates by   \citet{Cletal03}: (C/H)$_{\odot}=3.2\times
10^{-4}$ and (Si/H)$_{\odot}=4.0\times 10^{-5}$.\\
\addtocounter{table}{1}
\indent The properties of the grains along type A and B  anomalous curves, investigated in \citetalias{MB08}, show that B curves are characterized by a  number of small silicon  grains lower than normal curves with the same
R$_V$, and lower  than A curves too; they are also characterized by a  larger number of 
small carbon grains as A  curves are. Such results,  here revised accounting for recent updating of \citet{DL07},  are included in the figures to allow comparisons.
 The main change of such improvements (Table 2 of \citet{DL07}), is related to the  number of  carbon atoms per total H in each of the log-normal components (eq.s 11-14 of \citet{DL07}).
Appreciable effects occur when the parameters of grain size distribution provide a larger number of small carbon grain than the number required to fit normal lines,  thus in the case of anomalous lines.

 In the  figures, error bars of  WD normal curves span the range of values of different models corresponding to the same R$_V$ value (Section 3).

The results of our  WD best-fit models of type C curves are reported in Table \ref{resanom1}:  
the name of
sight line is in col. I, the dust-to-gas ratio in unit of 10$^{-2}$ in col. II, the
abundance ratios in col. III and IV, the 
small-to-large grain size ratios of carbon, R$_C$, in col. V and of silicon grains, R$_{Si}$, in col. VI,  the R$_V$ value in col. VII, and the 
E$_{(B-V)}$/N$_{H}$ ratio in col. VIII.
Here, as in the following, we define   small grains 
those with  size  $a\le 0.01\,\mu$m, and  large grains those with size above such a value.

\begin{figure}
\centering
\includegraphics[width=7.0cm]{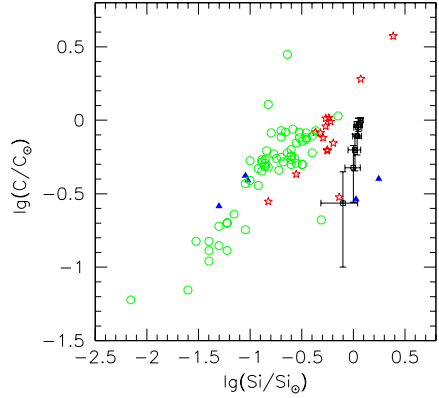}
 \caption{Abundances  of the dust locked 
up into the grains along our anomalous sight lines 
 compared with the solar values (see text);  (green) open circles  are for type C curves,
(black)  squares are for normal curves computed with WD parameters corresponding to seven different R$_V$ values,
 i.e. forty-nine models (Section 3). Also included are the results of type A  and B extinction models \citepalias{MB08},
(red) stars   and (blue) filled triangles respectively,   here revised accounting for latest updating \citep{DL07} (see text).
}
\label{ab_csi}
\end{figure}

\begin{figure}
\centering
\includegraphics[width=7.0cm]{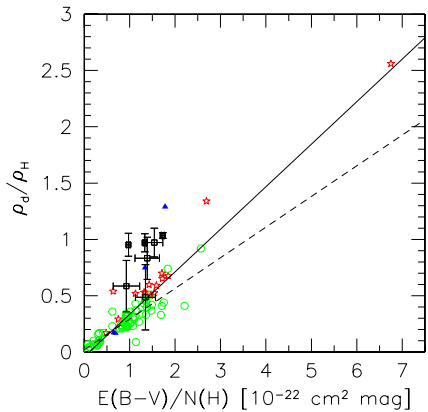}
 \caption{The behavior of the dust-to-gas ratio, $\rho_d/\rho_H$, normalized to the average
 Galactic value, 0.01, with the E(B-V)/N$_H$ ratio in unit of $1.7 \times 10^{-22}$
 mag\,cm$^2$ \citep{Boetal78};
symbols are as in the previous figure. The dashed line shows the
relation for type  C curves (eq. \ref{333}) and the continuous line that for all the
anomalous curves analyzed (eq. \ref{333_t}).
 }
\label{rdrh}
\end{figure}

The following conclusions can be derived by comparing the results presented in Table \ref{resanom1} with those derived  from the same dust grain models of normal extinction curves (Section 3):

i) Dust population models generally imply substantial abundances of elements
in grain material, approaching or exceeding the abundances believed to be
appropriate to interstellar matter \citep[and references therein]{D03}.
It is remarkable that, according to our models, the predicted amount of carbon  
that condenses into grains  along the majority of our  sight lines is 
lower than the average galactic value (Fig. \ref{ab_csi}). 
Only three  anomalous curves,  HD\,37061, HD\,21455,  and HD\,164492, require C/H abundance  larger than the solar value.
Moreover, for about 74\% of our models of type C curves, 
the predicted C/H ratio does not exceed a fraction  0.7-0.6 of C cosmic 
abundance which is accepted, although with large uncertainties, as the
average amount of carbon trapped in grains \citep{Dr09}.
No  anomalous line model exceeds the  solar value of the silicon abundance. 

ii) The dust-to-gas ratio of type C curves is linked to  the E(B-V)/N(H) ratio by the relation:
\begin{equation}
\frac{\rho_{d}}{\rho_{H}}=(0.2730\pm 0.019) \times \frac{E(B-V)}{N_{H}} +
(+0.020\pm 0.021)
\label{333} 
\end{equation}
with dispersion 0.076 (dashed line in Fig. \ref{rdrh}) and E(B-V)/N(H) in units of the average Galactic value, i.e., 1.7$\times 10^{-22}$ mag cm$^2$ \citep{Boetal78}. 
Accounting for all our anomalous curves we derive:
\begin{equation}
\frac{\rho_{d}}{\rho_{H}}=(0.377\pm 0.018) \times \frac{E(B-V)}{N_{H}} +
(-0.043\pm 0.025)
\label{333_t} 
\end{equation}
with dispersion 0.13, (continuous line in Fig. \ref{rdrh}),
in well agreement with the results of \citetalias{MB08}, although with a slightly
smaller correlation index, 0.92 instead of 0.95.

\begin{figure*}
\centering
\includegraphics[width=6.8cm]{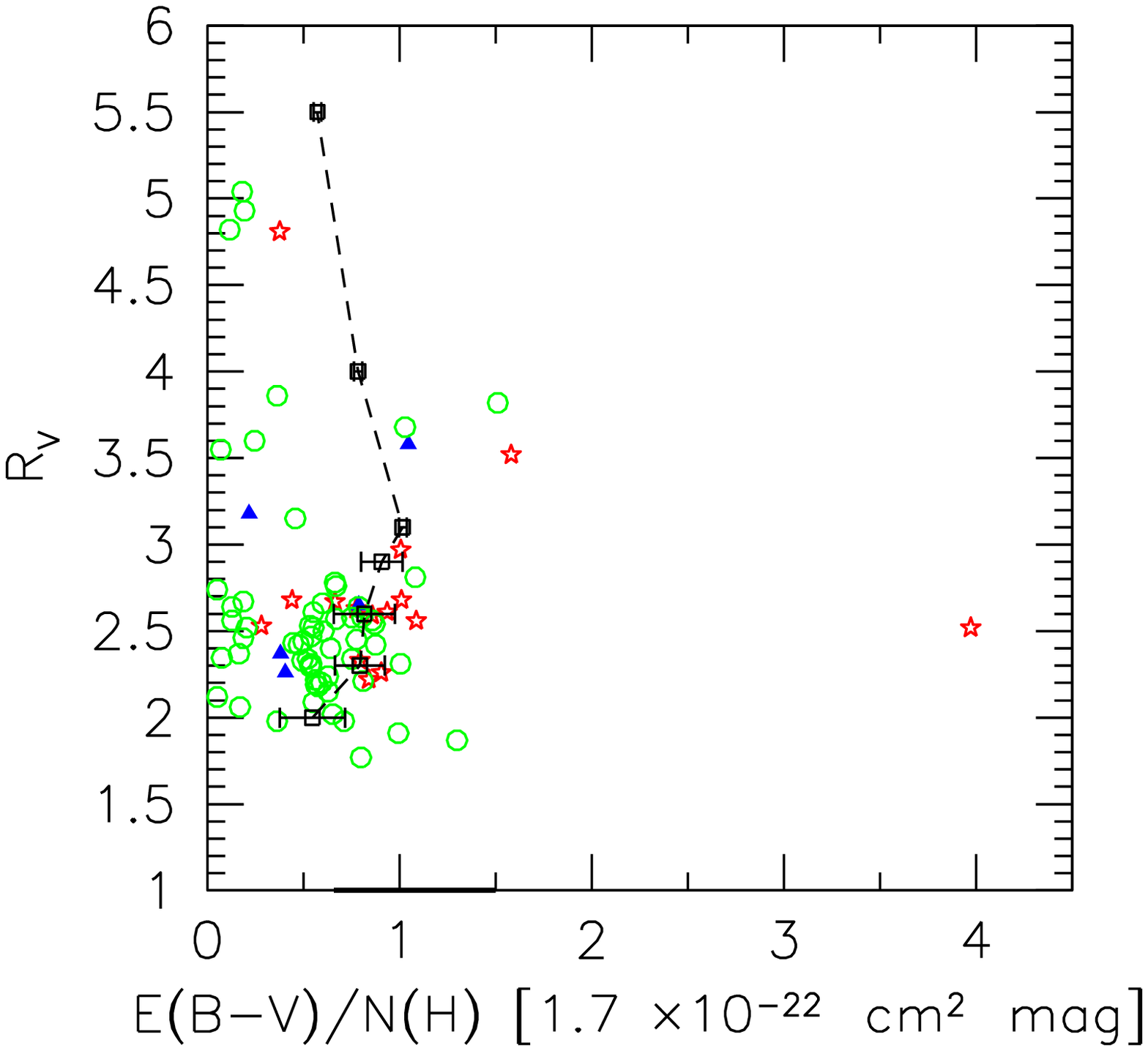}
\includegraphics[width=6.8cm]{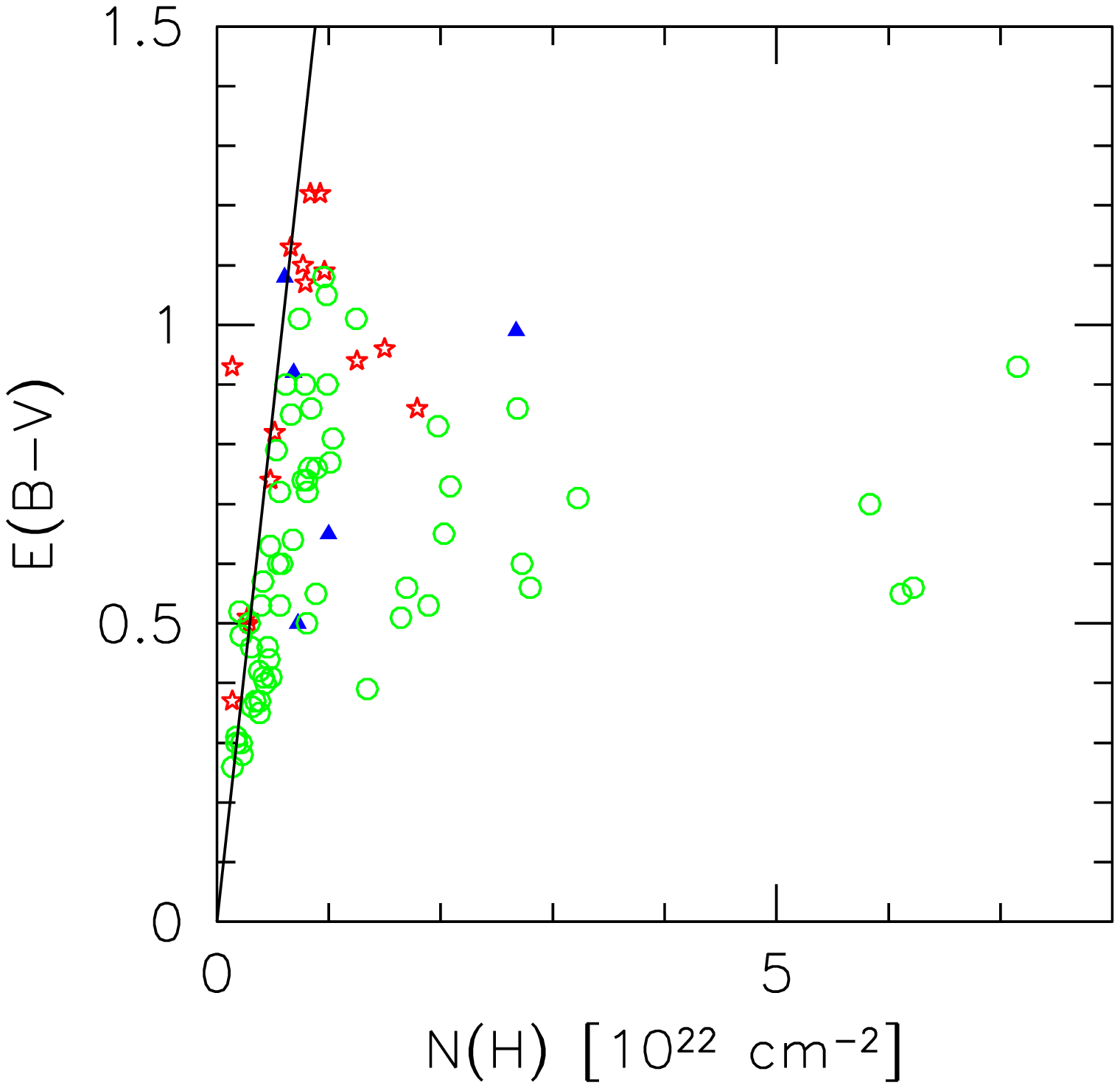}
 \caption { {\sl{Left panel}}: The R$_V$ values and the corresponding
  E(B-V)/N$_H$ ratios in unit of $1.7 \times 10^{-22}$
 mag\,cm$^2$, as derived from our WD dust grain models (see text); symbols are as in
 the previous figures;  the dashed line connects the results of   WD models of normal curves corresponding to seven different R$_V$ values 
(Section 3); the range in bold-face on x-axis shows the expected deviation of galactic curves \citep{Boetal78}.
{\sl {Right panel}}: The behavior of the reddening, E(B-V) \citep{Saetal85},  as a function of the predicted total hydrogen column density;
continuous line shows the galactic  average relationship found by analyzing a large sample of
sight lines whose HI and H$_2$ column densities were measured: E(B-V)/N(H)=1.7$\times 10^{-22}$ mag cm$^2$ \citep{Boetal78}; 
symbols are the same as in the previous figures.
 } 	
\label{rdrh1}
\end{figure*}

 Therefore,
anomalous curves are characterized by dust-to-gas ratios lower than the average galactic value.
In particular  no type C extinction curve exceeds the critical galactic value, 0.01  \citep[and references therein]{Baetal04}, independently
of its carbon grain abundance (see i)). 
Extinction curves of our sample having  E(B-V)/N(H) ratio equal to the average  Galactic value, exhibit their anomalous character due to a dust-to-gas ratio lower than  normal curves. 
 Moreover, their E(B-V)/N(H) can be affected by modifying the 
dust-to-gas ratio without any relevant change in R$_V$  unlike 
the behavior expected for  normal extinction curves (Fig. \ref{rdrh1}, left panel).

\begin{figure}
\centering
\includegraphics[width=7.0cm]{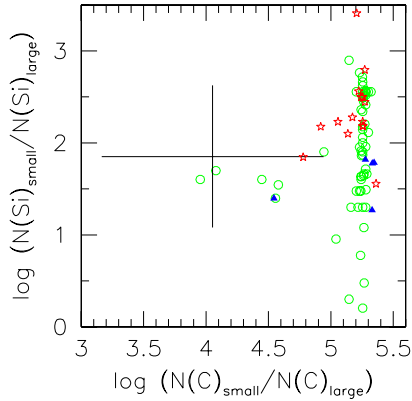}
 \caption{
The  predicted small-to-large grain size ratios for carbonaceous and silicate grains
of anomalous sight lines compared with the range spanned by WD models of CCM curves
(continuous lines) for the same range of R$_V$ values.
 }  
\label{ab_csi1}
\end{figure}

So, while anomalous curves can arise also in environments with normal reddening properties,
strong deviations  from the average reddening Galactic value  are a signature of anomaly as shown in 
Fig. \ref{rdrh1}, right panel, and discussed in \citet{Baetal04}. 
 
iii) In the large
majority of the cases, the expected small-to-large grain size ratios of anomalous lines
differ from the corresponding values expected for WD CCM  curves (Fig. \ref{ab_csi1}). Anomalous curves are generally characterized by a  number of small carbonaceous grains larger than normal curves  whereas  the small-to-large size ratios of  Si grains span a wider range,  from lower up to larger values than normal lines. 
Only five anomalous curves have normal values of both these  ratios, BD+59\,2829 and   BD+58\,310,  belonging to the 2$\delta$ sample, and  HD\,14707,  HD\,282622, BD+52\,3122, to the 3$\delta$ sample.
For these   sightlines  their best-fit  standard CCM curve in the UV range is below the observed data,
as outlined in Section 2 (Fig. \ref{anomB_sav}, right-panel).

\begin{figure*}
\centering
\includegraphics[width=6.8cm]{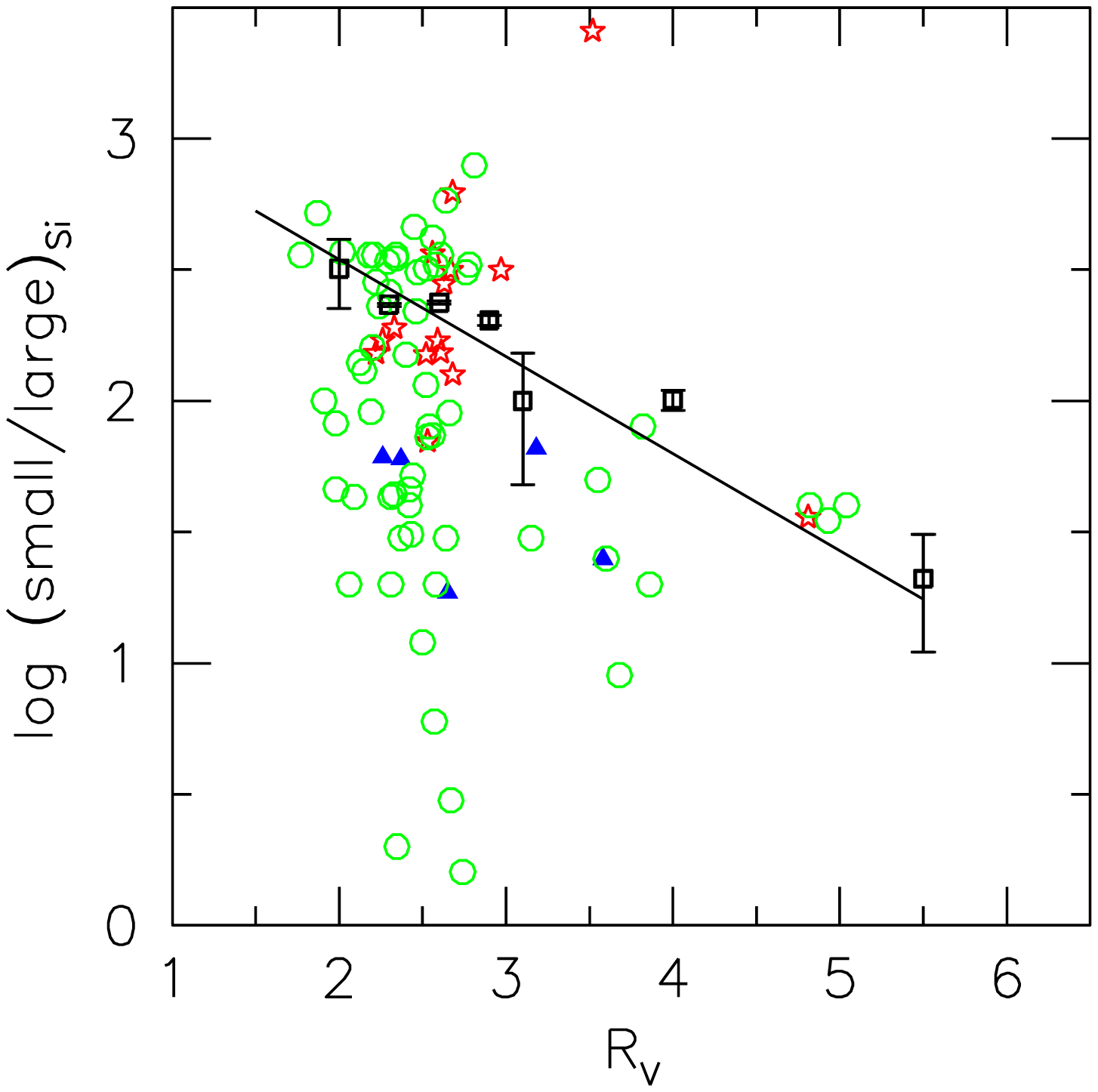}
\includegraphics[width=6.8cm]{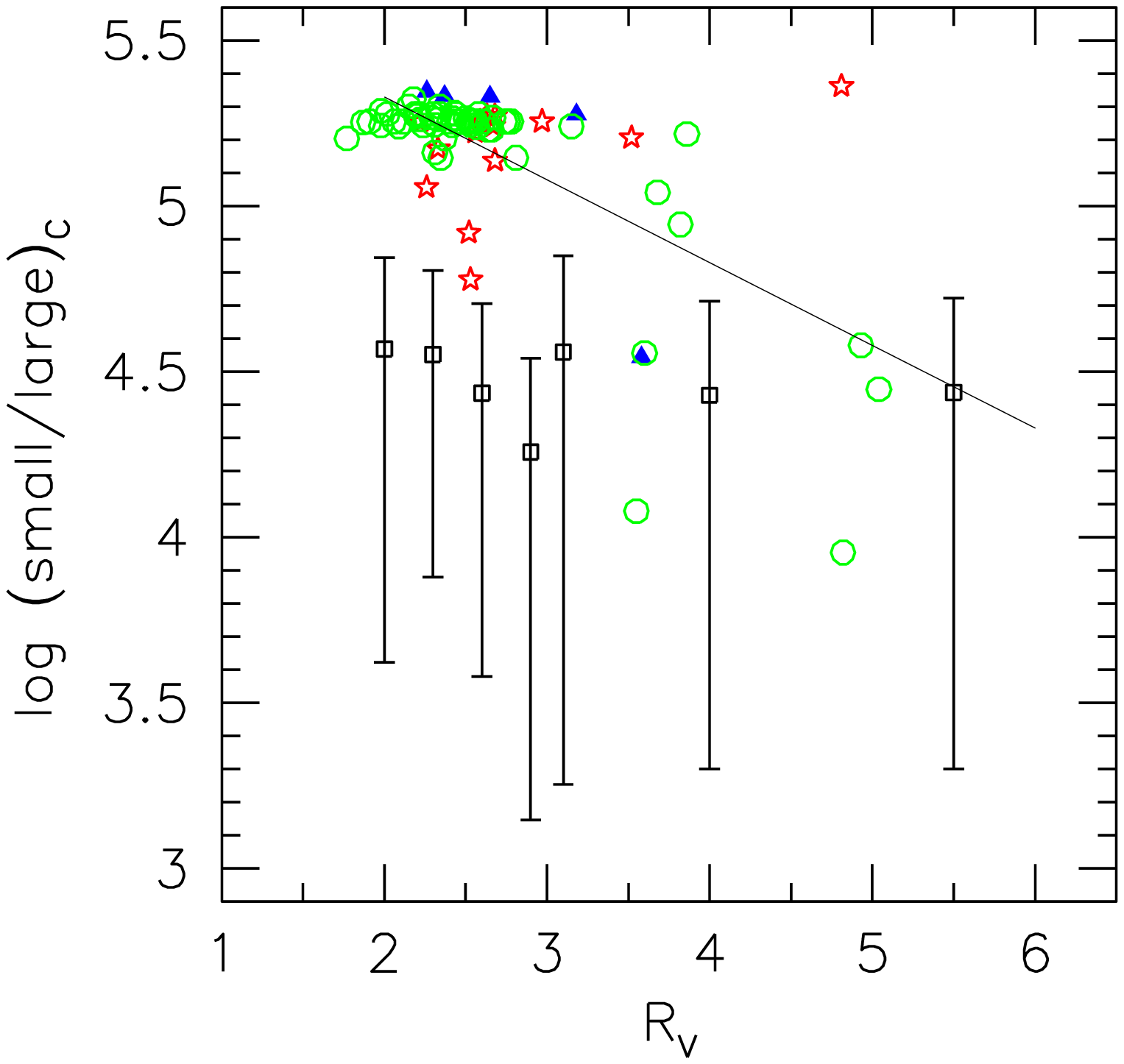}
 \caption{
The expected behavior of the small-to-large grain size ratio of silicate and carbonaceous dust
 against  R$_V$ values along anomalous sight lines
compared with the same along CCM curves
(open squares) computed with WD  parameters (Section 3).
Continuous line in the left panel corresponds to the relationship: y=(-0.37$\pm$0.03)+(3.30$\pm$0.09), with dispersion 0.18 and anti-correlation index -0.91, that applies to  normal lines.
Continuous line in the right panel shows  the relationship for the whole sample of anomalous curves: y=(-0.25$\pm$ 0.03)x $+$(  5.83$\pm$0.09) with  dispersion  0.19 and correlation index -0.65.}
\label{rgrainsi_rv}
\end{figure*}

\begin{figure*}
\centering
\includegraphics[width=6.8cm]{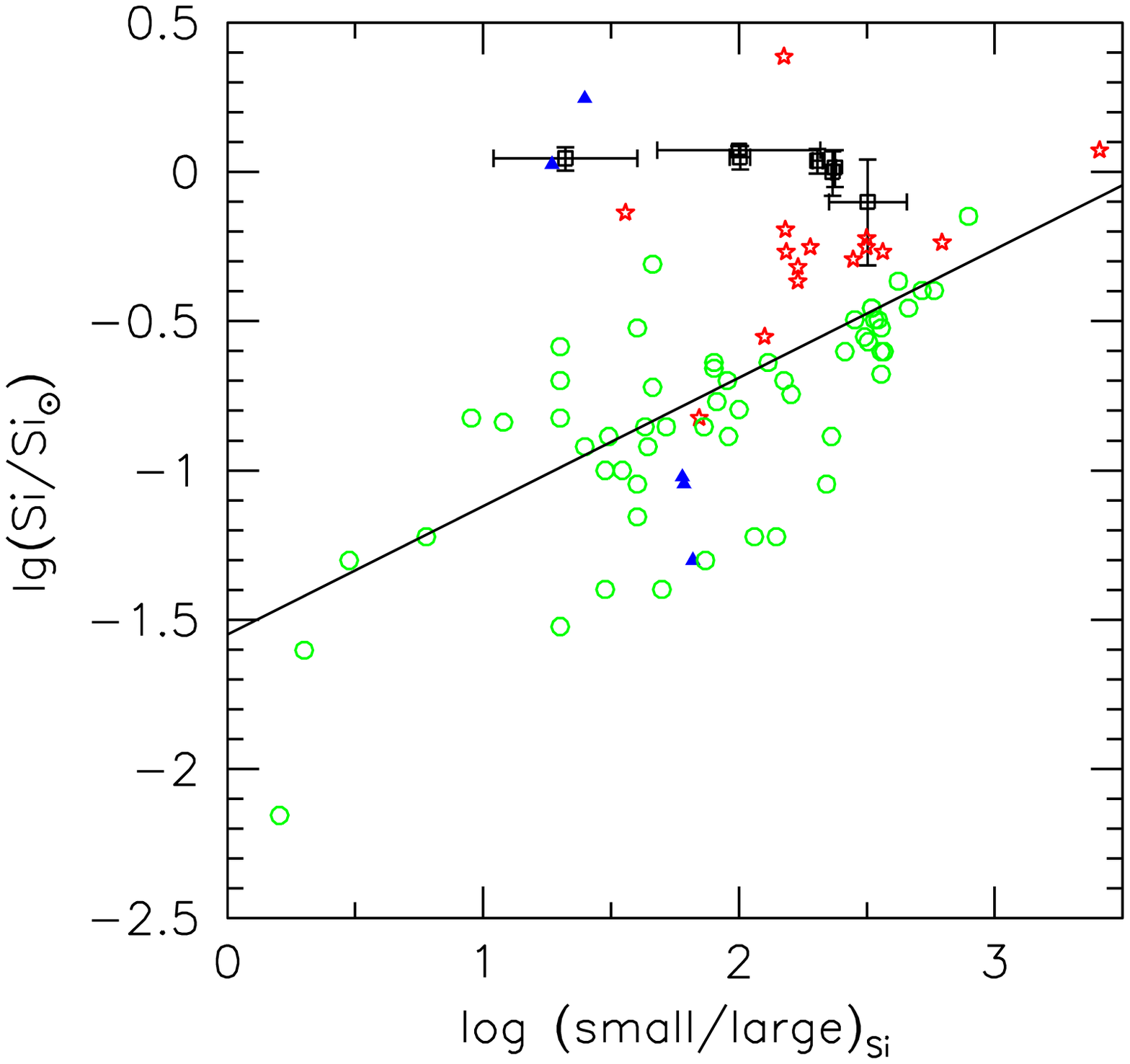}
\includegraphics[width=6.8cm]{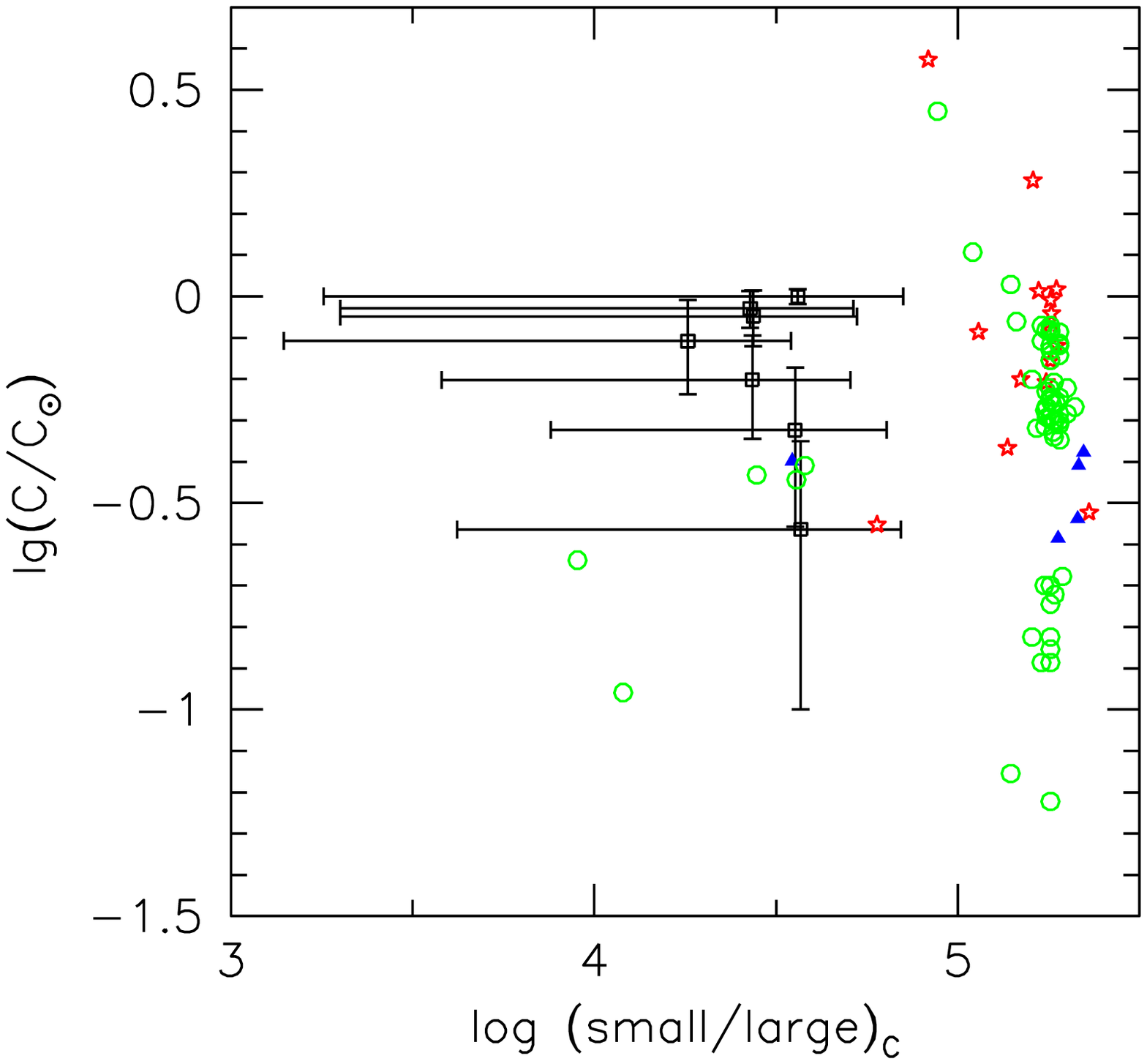}
 \caption{
The expected behavior of the small-to-large grain size ratio of silicate (left panel) and carbonaceous dust (right panel)
 against the silicon and carbonaceous dust abundances  of anomalous sight lines; continuous line holds for  Si grains of the whole sample: y=(0.43$\pm$0.06)x$+$(-1.55$\pm$0.13) with dispersion 0.34 and correlation index 0.60. Symbols are the same as in the previous figures. }
\label{rgrain_ab}
\end{figure*}

iv)  The  small-to-large size ratio of Si grains is almost independent of the selective extinction coefficient, R$_V$  (r=-0.20), 
 unlike the normal curves.  For  such curves this ratio anti-correlates strongly with R$_V$ (Fig. \ref{rgrainsi_rv}, left panel). The same ratio of carbon  grains for  type C curves is, indeed, anti-correlated with  R$_V$ with  anti-correlation index r=-0.80, at a variance with WD CCM lines  (Fig.\ref{rgrainsi_rv}, right panel).  Such anti-correlation is somewhat weakened by including all the anomalous curves in the sample.

v) From the analysis of Fig. \ref{rgrain_ab}, left panel, the small-to-large  size ratio  of Si grains along anomalous lines is  correlated with the Si abundance (continuous line in Fig. \ref{rgrain_ab}); type C curves are more correlated (r=0.73). The models predict  Si abundances up to hundred times lower than  solar values.

Compared with the characteristics of the environments where other types of lines of sight occur, those typical of C types   present, in average,  the  behavior summarized below and reported in Table \ref{res_sum}, where the mean values and their errors are in the same units as in Table \ref{resanom1};   N indicate WD normal curves:

1) The  small-to-large size ratio of carbonaceous grains is almost six times higher than the average value found for normal curves  and it is comparable to the  average value of  type A and B curves.

2) The small-to-large size ratio of silicon grains  is  smaller than the average value of A types and almost three times larger than that of type B curves, however it is like that of normal curves.

3) The average amount of carbon which condenses into grains  is  slightly lower than the expected galactic abundance trapped into grains, 0.7-0.6;  that is the value of  normal curves as derived here from the same grain models.
 Type A reaches the maximum value allowed; this is almost three times larger than that of B curves which show the lower value.

4)  All the anomalous curves show Si abundance lower than that of normal curves; type C, in particular, is characterized by the lowest amount.

Therefore, type C  sight lines require, in average,  dust abundances lower than the  abundances trapped into the grains along normal lines, in
particular of  silicate grains, as well as small-to-large grain size ratios   of carbonaceous dust larger
than expected for  normal curves.
 Such properties are different from those expected for A and B  curves. The former, in particular, correspond to sight lines with the highest  abundances of carbonaceous dust and the latter ones to  lines of sight with both the lowest carbon abundances and the lowest small-to-large grain size ratio of silicon dust, i.e., the largest Si grains. 

The different  properties of the dust locked into the grains along anomalous sight lines
can be recovered accounting for the violent nature of the interstellar medium. Shocks and grain-grain collisions should both destroy dust grains, so reducing the amount of the dust trapped in the grains,  and modify
the normal size distribution of the dust  increasing the small-to-large grain size ratio, as it  will be discussed  in the next section.

\begin{table}
\caption{Average properties of the dust locked up into the grains of WD extinction models}
\label{res_sum}
\centering
\begin{tabular}{llllll}
\hline\hline
T&$\frac{\rho_d}{\rho_H}$&$\frac{R_C}{10^{4}}$&$\frac{C}{C_{\odot}}$&$\frac{R_{Si}}{10^{2}}$&$\frac{Si}{Si_{\odot}}$\\
\hline
A&0.72$\pm$0.15  &15.8$\pm$1.13   &1.00$\pm$0.22 &3.79$\pm$1.61 &0.68$\pm$0.14\\
B&0.50$\pm$0.23  &17.5$\pm$3.54   &0.35$\pm$0.03 &0.46$\pm$0.10 &0.61$\pm$0.34\\
C&0.28$\pm$0.02  &16.4$\pm$0.59   &0.56$\pm$0.05 &1.69$\pm$0.22 &0.20$\pm$0.02\\
N&0.86$\pm$0.03  & 2.8$\pm$0.32   &0.71$\pm$0.03 &1.58$\pm$0.14 &1.04$\pm$0.03\\
\hline
\end{tabular}
\end{table}

\section{Discussion}


Table \ref{FMpar}   presents the values of the \citet{FM88, FM90} parameters (FM  parameters in the following)  of type C curves derived from our  models, since  a right estimate of such parameters from only five UV color excesses is impossible.

Then, the properties of  our sample can be compared with those of \citet{Vaetal04}, the larger and homogeneous sample of galactic extinction curves with known FM parameters and R$_V$ values available so far.
\citet{Vaetal04} found that the CCM extinction law,  with suitable R$_V$ values, applies for 93$\%$ of their 417 sight lines  and that only four lines deviate by more than 3$\sigma$. They conclude  that the physical processes that give rise to grain populations that have CCM-like exctinction dominate the interstellar medium.

Sixteen of curves here have been studied also by  \citet{Vaetal04}, five belonging to the 3$\delta$  sample,
HD\,14357,  HD\,37061,  HD\,164492,  HD\,191396,  BD$+$57\,252, and eleven to the 2$\delta$ sample,
 HD\,54439, HD\,96042, HD\,141318, HD\,149452, HD\,152245,  HD\,168137, HD\,248893,  HD\,252325,  BD$+$59\,2829, BD$+$62\,2154, and BD$+$63\,1964. 
By comparing  their parameterized UV extinction curves at the five ANS wavelengths with our ones, we find meaningful differences, i.e., larger than three $\sigma_r$
at one wavelength or more, for all the common curves of the 3$\delta$ sample, unless for  BD$+$57\,252 which well agrees with our data.
Concerning the sight lines belonging to  the 2$\delta$ sample,  three curves (i.e., HD\,168137 and HD\,252325, and HD149452)
show differences larger than two $\sigma_r$ at four  wavelengths, one curve (i.e., HD\,248893)
 at two wavelengths  and  four curves (i.e., HD\,152245,  BD$+$59\,2829, BD$+$62\,2154, and BD$+$63\,1964) at one wavelength.
HD\,96042 well agrees with our data, and the remaining curves, i.e., HD\,54439 and  HD\,141318, show only 
differences lower than two $\sigma_r$ at one wavelength.

 It must be remarked, however, that  spectral type and luminosity class  of \citet{Vaetal04} are based on spectral properties in the UV  rather than in the visible spectral range as in \citet{Saetal85};
moreover the color excesses used by \citet{Vaetal04} are derived from IUE ({\it International Ultraviolet Explorer}) spectra using the pair method, whereas \citet{Saetal85} 
used the  ANS photometry and  intrinsic colors by \citet{Wuetal80} as described in Section 2.

\begin{figure}
\centering
\includegraphics[width=10.cm]{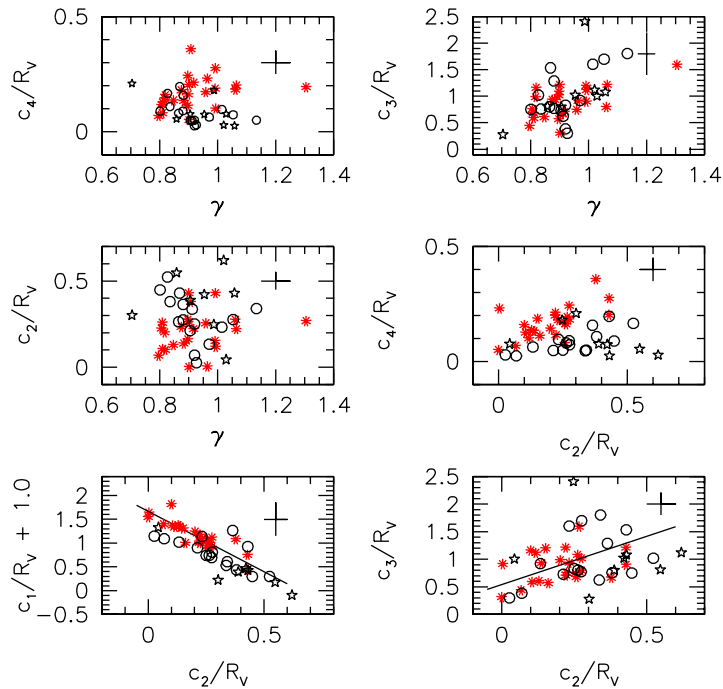}
 \caption{
 Open circles and stars  (black) show  the FM parameters of sixteen
 extinction curves derived from our best-fit dust models in Section 3  together with those of eight curves in Paper 1, respectively;
 asterisks (red) are  for the same curves by
\citet{Vaetal04}; continuous lines show the relations
 as derived by  \citet{Vaetal04} (see their Table 6).
}
\label{common_par}
\end{figure}

Fig \ref{common_par} compares the  FM parameters of sixtheen common sight lines here  together with those (eight) in Paper 1.


\subsection{The FM parameterization }
\begin{figure}
\centering
\includegraphics[width=10.cm]{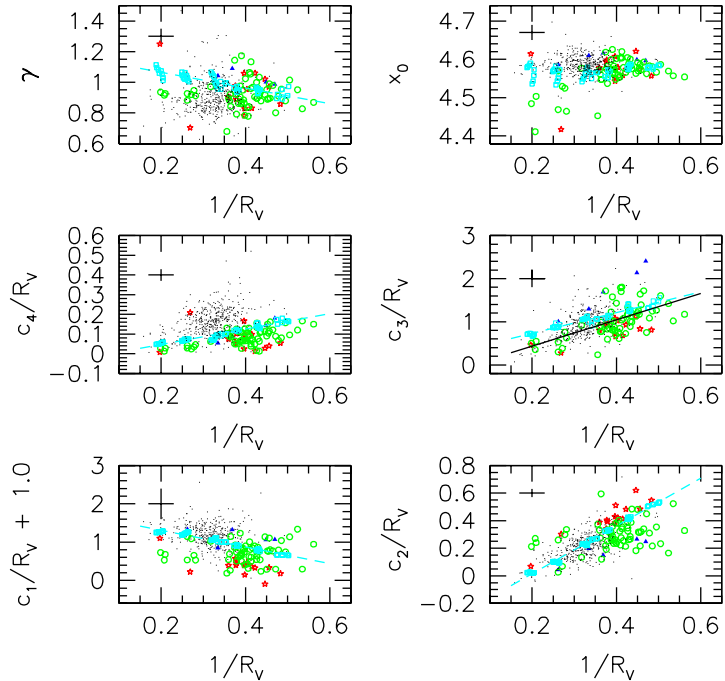}
 \caption{Black points show the FM parameters of
 extinction curves in  \citet{Vaetal04}; open circles (green) are the FM parameters of  C type anomalous curves as  derived from our models (see Sect. 3), (red) stars and (blue)
 triangles are the same of A and B types, and (cyan) open squares  of WD CCM models (see text); black continuous lines are the relationships in Table \ref{rel_par} with r$>$0.50, (cyan) dashed lines are the same for WD CCM curves (Table  \ref{rel_par_N}).}
\label{par}
\end{figure}

\begin{figure}
\centering
\includegraphics[width=9.5cm]{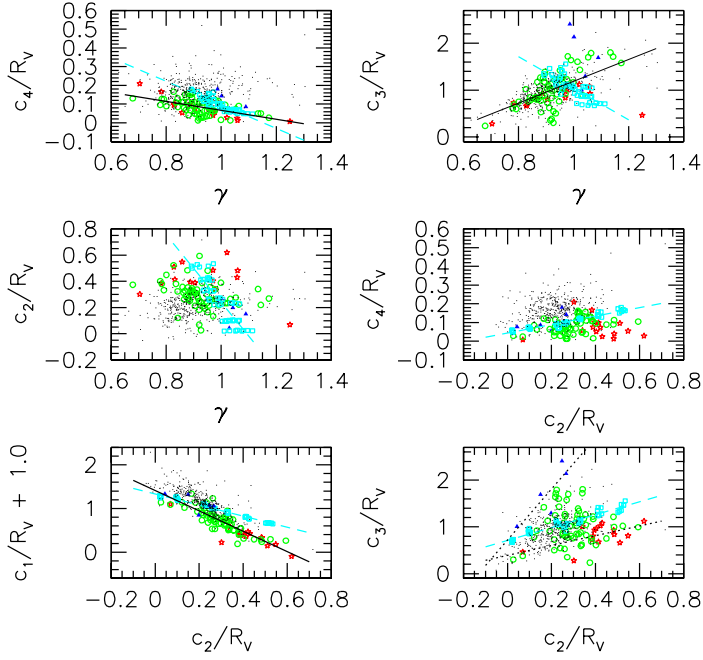}
 \caption{FM parameters of our extinction curves compared with the sample of
 \citet{Vaetal04}; symbols are as in the previous figure; 
 dotted lines in the  bottom right panel are correlations for A and B types \citepalias{MB08}.
 }
\label{para}
\end{figure}

\begin{figure}
\centering
\includegraphics[width=9.5cm]{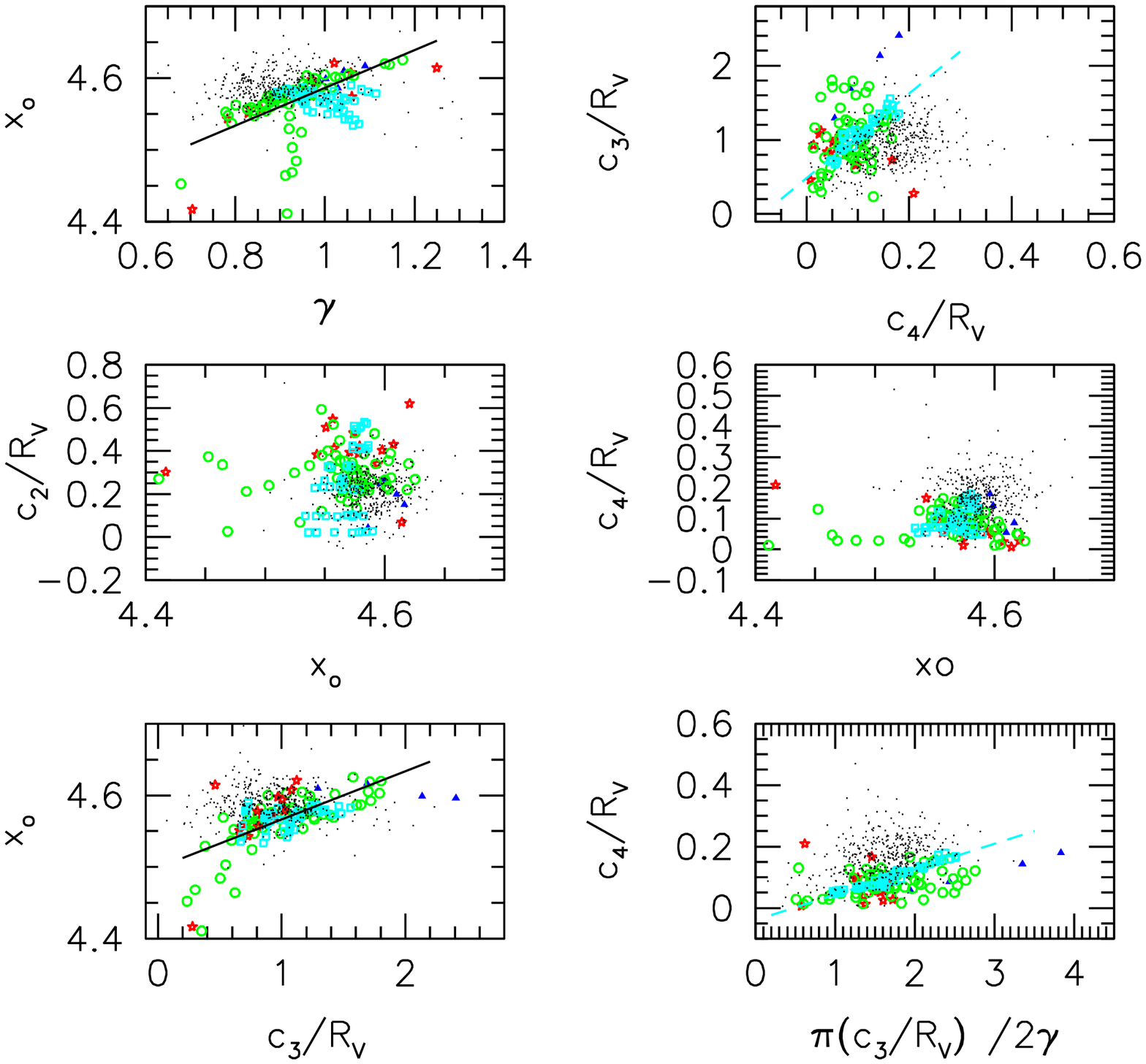}
 \caption{FM parameters of all our anomalous curves compared with the sample of
 \citet{Vaetal04}; symbols are as in Fig. \ref{par}; dotted line is for B types and dashed line for  A types, following Table 5 in \citetalias{MB08}.
 }
\label{para1}
\end{figure}

In the following we 
define $c'_i=c_i/R_V$ for i=2,3, and 4,  $c'_1=c_1/R_V + 1$, $x_o$ and $\gamma$
being the same, to compare our results  with
those of \citet{Vaetal04}. The data in Table  \ref{FMpar}
concern the parameterization in terms of the normalized color excesses,
i.e., $\kappa(\lambda)$=E($\lambda$-V)/E(B-V),
whereas the primed symbols, i.e., $c'_i$, are in term of  A($\lambda$)/A$_V$.
Fig.s \ref{par},  \ref{para}, and Fig. \ref{para1},  compare the  FM parameters of all our anomalous curves
with  both the FM parameters derived from WD CCM curves (Section 3), and  the sample of  \citet[their Table 5]{Vaetal04}.

Figures \ref{par}, \ref{para}, and Fig. \ref{para1} show 
the  relations between the different  parameters.  Relationships with (anti)/co-relation index, r,  larger than (-)0.4  are reported in Table  \ref{rel_par_N} for WD normal CCM curves, in Table  \ref{rel_par_C} for C types,  and in Table \ref{rel_par} for all the anomalous sight lines.

 Looking at Fig.s \ref{par}, \ref{para}, and Fig. \ref{para1}  important  agreements with the results of \citet{Vaetal04} are emphasized as well as similar trends as WD normal curves. These points show that dust models of anomalous curves, which best-fit ANS data, are not biased by the low resolution of such data.
 In particular:

\begin{table}
\caption{Relations between FM parameters of WD CCM curves }
\label{rel_par_N}
\centering
\begin{tabular}{lll}
\hline\hline
 r &          linear fit& dispersion\\
\hline
$-$0.98& $c'_1=(-2.177\pm0.264)/R{_V}+(1.746\pm0.022)$ &0.0428\\
+0.99& $c'_2=(1.731\pm0.020)/R{_V}-(0.331\pm0.007)$ &0.013\\
+0.97& $c'_3=(2.409\pm0.092)/R{_V}-(0.246\pm0.032)$ &0.060\\
+0.96& $c'_4=(0.394\pm0.016)/R{_V}+(0.032\pm0.006)$ &0.010\\
$-$0.87& $\gamma=(-0.508\pm0.041)/R{_V}+(1.166\pm0.014)$ &0.027\\
$-$0.99& $c'_1=(-1.269\pm0.024)c'_2+(1.333\pm0.007)$& 0.028\\
$-$0.90& $c'_4=(-0.628\pm0.045)\gamma+(0.726\pm0.045)$ &0.017\\
+0.96 & $c'_3=(1.375\pm0.057)c'_2+(0.711\pm 0.017)$&0.065\\
+0.96& $c'_4=(0.226\pm0.009)c'_2+(0.044\pm0.003)$ &0.010\\
-0.86& $c'_2=(-2.577\pm0.216)\gamma+(2.816\pm0.216)$ &0.010\\
-0.79 & $c'_3=(-3.380\pm0.374)\gamma+(4.420\pm 0.374)$&0.143\\
+0.93 & $c'_3=(5.664\pm0.321)c'_4+(0.488\pm0.034)$&0.086\\
+0.96& $c'_4=(0.083\pm0.003)c'_3\pi/2\gamma^2+(0.040\pm0.006)$ &0.011\\

\hline
\hline
\end{tabular}
\end{table}

\begin{table}
\caption{Relations  between FM parameters of type C anomalous curves}
\label{rel_par_C}
\centering
\begin{tabular}{lll}
\hline\hline
 r &          linear fit& dispersion\\
\hline
+0.61& $c'_3=(3.086\pm0.511)/R{_V}+(-0.209\pm0.208)$ &0.300\\
+0.61& $c'_4=(0.330\pm0.054)/R{_V}-(0.047\pm0.022)$ &0.032\\
+0.60& $x_o=(0.320\pm0.054)/R{_V}+(4.432\pm0.022)$& 0.032\\
$-$0.83& $c'_1=(-2.219\pm0.186)c'_2+(1.379\pm0.058)$& 0.144\\
$-$0.48& $c'_4=(-0.224\pm0.053)\gamma+(0.297\pm0.049)$ &0.035\\
+0.43& $c'_4=(0.173\pm0.047)c'_2+(0.034\pm0.015)$ &0.036\\
+0.53& $x_o=(0.246\pm0.048)\gamma+(4.338\pm 0.045)$&0.034\\
+0.68 & $c'_3=(3.021\pm0.417)\gamma-(1.767\pm 0.386)$&0.278\\
+0.76&   $x_o=(+0.080\pm0.009)c'_3    +(4.483\pm0.010)$&0.026\\
\hline
\end{tabular}
\end{table}

\begin{table}
\caption{Relations between FM parameters of all the anomalous curves}
\label{rel_par}
\centering
\begin{tabular}{lll}
\hline\hline
 r &          linear fit& dispersion\\
\hline
+0.55& $c'_3=(3.064\pm0.515)/R{_V}+(-0.183\pm0.207)$ &0.344\\
+0.44& $c'_4=(0.262\pm0.059)/R{_V}-(0.019\pm0.024)$ &0.040\\
+0.48& $x_o=(0.262\pm0.054/R{_V}+(4.464\pm0.022)$& 0.036\\
$-$0.88& $c'_1=(-2.326\pm0.142)c'_2  +(1.405\pm0.047)$& 0.149\\
$-$0.54& $c'_4=(-0.248\pm0.043)\gamma +(0.315\pm0.040)$ &0.037\\
+0.62& $x_o=(0.263\pm0.037)\gamma+(4.322\pm 0.035)$&0.032\\
+0.55 & $c'_3=(2.331\pm0.399)\gamma-(1.145\pm 0.374)$&0.346\\
+0.67&   $x_o=(+0.067\pm0.008)c'_3  +(4.499\pm0.009)$&0.030\\
\hline
\hline
\end{tabular}
\end{table}

i) The correlation  between c$'_3$, the bump height, and 1/R$_V$ for the whole sample exhibits
the same slope as both WD normal curves (Table  \ref{rel_par_N}) and the sample of \citet{Vaetal04}, that shows the same correlation index as our sample  (slope 3.48$\pm$0.24, and dispersion 0.25).

Indeed each type of anomalous curve obeys a different relation:
B  curves show the stronger correlation index (r=0.99) and the steeper slope (6.77$\pm$0.46); A curves have an intermediate correlation index, r=0.64, but the lower slope  2.10$\pm$0.64.

ii) $\Gamma$, the full width at half maximum (FWHM) of the bump, spans a large range of values by changing 1/R$_V$   as the sample of  \citet{Vaetal04},  perhaps reflecting a wide range of environments  \citep{CC91}.  No correlation is found for our sample as well as for each anomalous type, unlike WD CCM models.

iii) For the whole sample of anomalous curves,  c$'_4$, the  far-UV (FUV) non linear rise, and 1/R$_V$  correlate  with  a lower correlation index  than  WD normal curves  but almost with the same slope;
B types are better correlated  (r=0.87) than C types, whereas  A types do not correlate (r=-0.20). The sample of  \citet{Vaetal04} shows a weaker correlation, r=0.38, and  a higher  slope, 0.51$\pm$0.06, than our findings for the total sample.

iv) The  correlation between $c'_2$ and 
$c'_1$  (Fig. \ref{para}) shows how tightly constrained are the linear components of the extinction.
The whole sample of  anomalous curves is more strongly correlated  (Table \ref{rel_par}) than type C curves (Table \ref{rel_par_C}). Their slope   agrees within  the errors with the findings of  \citet[their Table 6]{Vaetal04} but  it is  steeper  than that derived from WD CCM models, more than three times the error.
 As discussed in the previous section, this difference is a consequence of the lower  amount of dust grains with normal and large sizes  which affects the optical portion of the extinction curve of  anomalous sight lines compared to normal lines.

v) The parameters $c'_4$ and $c'_2$ do not correlate (r=0.17) for the whole sample of anomalous curves, 
in agreement with the findings of \citet[and references therein]{Vaetal04}. Thus the carriers of the FUV non-linear rise are not the same as the optical linear rise.  For WD CCM curves such parameters, indeed, are correlated.   A  weak correlation arises  for C types  (Table \ref{rel_par_C}).

vi)  Our results between $c'_2$ and $\gamma$ (r=-0.31) are in agreement with those
\citet[and references therein]{Vaetal04} which do not find any correlation. 
 For WD CCM models such parameters are  anti-correlated, wider bumps are found in extinction curves with  weaker  linear rises.

vii) For the whole sample of anomalous curves   we  find the same correlation  between  $c'_3$ and $\gamma$
as that of \citet{Vaetal04}  (Table \ref{rel_par}), with a correlation index  slightly lower than   their (their Table 6, r=0.58). It means that as the bump FWHM increases, the bump strength is also increasing.  C types are  better correlated and with a steeper slope than the whole sample here. \citet{JG93} attributed this relation, at least  partly,  to  the fitting procedure whereas,
following \citet{FM88}, such parameters are truly related in some way. 
WD CCM curves span a  shorter range of $\gamma$ and  $c'_3$  values which are anti-correlated. 

viii) There is almost  no correlation between $c'_4$ and $c'_3$ for both the whole sample (r=0.24) and  type C curves (r=0.25),
as well as for the sample of \citet{Vaetal04} (r=0.31), whereas the correlation is strong for WD CCM curves.
As discussed in \citetalias{MB08}, A types are weakly  anti-correlated (r=-0.56) and B types are strongly correlated (r=0.92).

Therefore,  looking at the same figures, several important differences arise from our sample and that of \citet{Vaetal04}.

i) There is a weak relation  between   $x_o$, the bump position ($\mu m^{-1}$), and 1/R$_V$ that is stronger for type C curves, at a variance with the results both of \citet{Vaetal04} and of WD CCM models (r=0.31).

ii) For the whole sample of  anomalous curves $c'_4$ and $\gamma$ are
 anti-correlated in the sense that a broader bump is found along sight lines  with smaller FUV non linear rise;  by considering  only  type A curves, the anti-correlation index increases (r=-0.87)  whereas, by considering only C types,  the anti-correlation index decreases (Table \ref{rel_par_C}).
No correlation is found by  \citet{Vaetal04} while \citet{Ca86}, \citet{FM88},
and \citet{JG93}  obtained an opposite trend, in the sense that 
a wider bump is found along sight lines  with larger FUV rise.
It must be remarked  that WD CCM curves show a very similar trend as anomalous curves  although with a steeper slope  and a higher  anti-corelation index.
 This finding cannot be ascribed to the less well sampling in wavelength of ANS data since a similar trend arises from dust models fitting also complete extinction curves.
This is provided by the different  dust components that play the job in our models.

iii) The parameters $c'_2$ and $c'_3$ do not correlate
for the whole sample of anomalous curves (r=-0.05) as well as for C types (r=0.06),  at 
variance with the results of \citet[their Table 6, r=0.49]{Vaetal04}  and with those of WD normal curves.
However, when considering separately type A and B curves 
good correlations are found though  with  very different slopes \citepalias{MB08}.
In particular, looking at Fig.  \ref{para},
A and B   curves outline  the  lower and upper limits of the region where C curves mix to normal ones.
The correlation occurring separately for 
A and B types  suggests that   some fraction of the linear rise
is associated with the bump \citep{Ca86} but in a different proportion for such types, as discussed in \citetalias{MB08}.
C type curves,  characterized by intermediate properties of their grain populations compared with A and B types, as outlined in the previous section,  have no the same proportion of grains which contribute to the bump and to the linear rise, thus do not  correlate.

So, anomalous curves show bump properties, i.e., bump width, bump  height,  bump strength, $c'_3/\gamma^2$, and   bump position, well correlated whereas the same properties are independent of the  linear rise  $c'_2$  (Fig. \ref{para1}).  
Neither WD CCM models, neither the sample of  \citet{Vaetal04}  show any correlation with the bump position.
 Moreover, the bump height, $c'_3$  correlates with the linear rise, $c'_2$,  both for the sample of  \citet{Vaetal04}  and for WD CCM models  (Table \ref{rel_par_N}), though with different slopes,
 showing that bump properties are driven by different dust components, contrary to what happens for anomalous curves.

WD models of normal curves show a good correlation between the bump area, $\pi c'_3/(2\gamma)$, and the FUV non-linear rise, emphasizing that the same grain populations concur to these features, whereas  C type curves show a weak correlation which weakens further by considering the whole sample of anomalous curves (r=0.40), since  A types do not correlate (r=-0.28).

 Therefore, several extinction properties of type  C  anomalous curves differ from those of  CCM curves computed with the same dust models and corresponding to the same R$_V$ values, showing  that mechanisms  working in the environments of anomalous curves are different from those in the environments of normal curves.

\subsection{Insight into environmental conditions}

The results here derived from    dust grain models  \citep{WD01,DL07},
show that the dust-to-gas ratios predicted for our sample are lower than the average galactic value (Section 4). 
For A types, in particular, the  average value of such a ratio is the larger value, whereas for C types is the lower one.
Such a trend is a consequence of the under abundance of Si grains, which,  for A types, is in average 1.5 times less than that of normal, CCM curves computed with the same grain models. However it is at least  three times greater than that of type C  curves (Table \ref{res_sum}). 
Moreover, the anomalous behavior of type C curves is driven  by a larger population of small grains than the normal population, i.e., that characterizing lines of sight with extinction properties like  CCM  curves.

Many theoretical studies have faced the problem of the influence of shock waves on 
grains and their size distributions (\citet{{Jones091},{Jones092}} and \citet{Dr09} for a review).
As much as 5\%-15\% of the initial grain mass (a $\ge 0.005\,\mu$m) may be
end up in very small fragments with a$\simeq .0014\,\mu$m
in shock waves expanding in a  warm interstellar medium with shock velocities
 between 50 and  200\,km/s \citep{Joetal96}.
High velocity shocks affect grains through sputtering reducing the number of 
small particles, while in
shocks with lower velocities grain-grain collisions alter the size 
distribution by increasing the small-to-large grain size ratio \citep{Jo05}. 
Following \citet{C78}, silicate grains may be almost
destroyed by velocity shock higher than $\simeq$80\,Km/s while graphite grains
require velocities higher than 100\,Km/s.
\citet{Joetal96} found that, for a  shock velocity of 100\,Km/s, the percentage of the initial mass of silicate
grains destroyed increases from 18\% to 37\% by increasing the average density of the pre-shock
gas, $n_0$,  from 0.25 to
25\,cm$^{-3}$; for the same conditions, that of destroyed carbonaceous grains increases from
7\% to 13\%. Detailed description of the various grain destruction mechanisms and grain lifetime in the 
interstellar medium were presented by \citet{Joetal96} and \citet{Jo04}.
Moreover, with shocks having  high 
velocities, the destruction of silicate grains by sputtering can reduce the 
depletion of Si \citep{BS77}.  Recently, \citet{Guetal09} found that silicon dust is destroyed in J-type shocks slower than 50\,Km/s by vaporisation not sputtering.

In order to gather information on the physical nature and the behavior of grains, the knowledge of the environments crossed by the sightlines is as much essential as the shape of the extinction curves. Unfortunately the knowledge of the environmental properties is advancing slowly, since various relevant data are still not available for many sightlines, as the column density of the different gas constituents and the depletion of the most important chemical elements.
However,  for several anomalous lines of sight analyzed in \citetalias{MB08},  A and B types, there
is convincing evidence that their environments have been processed by shock
waves.  Similar conditions are reported in the literature for
some lines of sight  of C type here analyzed,  as summarized  in the following.

Multi-object spectroscopy  toward {\sl h} e {\sl $\chi$} Persei open clusters \citep{Points04} revealed the great complexity of the interstellar Na I 
absorption in the Perseus arm gas. Velocities from -75 down to -20 Km/s, characterize  such region.
The  intermediate velocity (-50 Km/s) component revealed in the south region of $\chi$ Persei,  where HD\,14357 belongs,
 corresponds to an intervening interstellar cloud \citep{Points04}.

HD\,37061 is a translucent  sight line \citep{vanD88} which crosses M43, an  apparently spherical HII region ionized by its star, HD\,37061.

The triple star  HD\,28446 (DL Cam), with its H$\alpha$ emission region, is located near the top of a ring of dust and small dark clouds in the Cam OB1 layer \citep{StraLa09}.

The sight line  HD\,73420  crosses the Vela OB1 association including the Vela X-1 binary pulsar system \citep{ReeCam01}. 

The line of sight  HD\,152245   crosses a bright  HII region, RCW 113/116, associated to several isolated molecular clouds located at the edge of evolved HII regions. They are thought to result from the fragmentation of the dense layer of material swept up by the expanding HII region \citep{Uretal09}.

\citet{Geor96} report H$\alpha$ and CO velocities respectively of -20 and -25 Km/s in NGC6193, where  HD\,149452
belongs.

In the  nebula Simeiz 55 where HD\, 191396 belongs,
high-velocity motions are reported \citep{Esi96}.

 HD\,248893 belongs to the Crab Nebula which is a well known supernova remnant \citep{wu81}.

 HD\,252325 sight of line crosses the compact HII region/molecular cloud complex G189.876$+$0.156 squeezed by the stellar wind from massive stars  \citep{Quinetal08}. 

 HD\,253327 is one of the ionizing stars of the compact HII region/molecular cloud complex G192.584$-$0.041  where a stellar wind is sweeping up the surrounding material \citep{Quinetal08}.

The young open cluster NGC\,6823, where  HD\,344784 and  BD$+$23\,3762 belong, at the edge of the  Vulpecula rift molecular cloud, arises in a region where star formation is probably triggered by external shocks \citep{Fre99}.

The sight line  BD$+$62\,338 crosses the most notable Galactic-plane  high velocity cloud complex, named complex  H by \citet{WaeWo91}, where the velocity of the  neutral hydrogen cloud relative to the Local Standard of Rest is -201 Km/s  \citep{Wa01}.

\section{Conclusions}

With aim at deepening our knowledge into the extinction properties,
more than sixty  extinction curves singled out from the same sample as defined  in \citetalias{MB08} have been analyzed.
In that paper 785 UV extinction curves from the ANS catalog and IR data from 2MASS catalog (Section 2) were compared with   standard CCM  curves for a variety of R$_V$ values in the range 2-6. The curves were classified as normal if they fit at least one of the CCM curves or anomalous otherwise. 
Eighty-four curves were retained which deviate by more than two $\sigma$
from their  standard CCM best-fit law at least at one UV wavelength (eq. 2).
In this paper sixty-four  anomalous sight lines, defined as type C curves (Section 2),  have been examined. 
For the majority of such lines  the corresponding best-fit CCM curve is always well above  the UV   data ($\ge$ 2$\sigma$ at least one UV wavelength; Fig. \ref{anomB_sav}, left-panel).
In  \citetalias{MB08} the extinction properties of twenty anomalous curves of different types, A and B type, were studied.  Type A curves are characterized by weaker bumps and steeper  far-UV rises  than
expected from their standard best-fit CCM curve, that is worse, of course, by more than 2$\sigma$ at least  one UV wavelength for each of them; type B  curves show stronger bumps  together with smoother far-UV rises.

By fitting the observed  curves  with extinction curves provided by dust grain models, we aim at giving insight into
the properties of the grains,  the processes  affecting them, and  their relations with the environmental
characteristics along selected lines of sight.

 The selected sightlines represent the larger and homogeneous sample of extinction curves with large UV deviations from CCM law studied so far with dust models.

The  models are reckoned by following the 
prescriptions of \citet{WD01} i.e., using their grain size distributions 
together with the more recent updating \citep{DL07}, as in \citetalias{MB08}. 
Models of \citet{WD01} are able to reproduce the observed wavelength-dependent extinction law  of normal curves in the local MW for R$_V$ values 3.1, 4.0, and 5.5, in the Large Magellanic Cloud, and in the Small Magellanic Cloud (SMC, bar region); moreover such models are also consistent with the observed IR emission from diffuse clouds in the MW and  in the SMC \citep{LD02}, and in several nearby  galaxies \citep[and references therein]{Liu10}.
This choice, the same as in \citetalias{MB08}, allows us to
compare the results of our deviating curves with the same  as normal curves in a self-consistent framework, and thus to recover the relative trends of the dust properties along selected sight lines  overcoming the modeling uncertainties, widely discussed by \citet{Dr09}.
 Since our anomalous sample extends to small R$_V$ values, down to 2.0, twenty-four dust models of CCM curves with  R$_V$ smaller than 3.1  are built up with the same grain models to allow the comparison.

The results derived from  models, both of CCM curves  and  of anomalous lines of sight, are presented in terms of dust-to-gas ratios, abundance ratios, small-to-large grain size ratios of the dust locked up into the grains  following the recipes in Section 4.
 Results of \citetalias{MB08} are also revised to account for recent updating \citep{DL07} (Section 3 and 4).
 Moreover,  FM parameterization 
 of all the dust models has been also performed in order to compare the results with the  sample of  \citet{Vaetal04}.
 Anomalous curves show bump properties i.e., bump width, bump strength, bump height, and the bump position well correlated   whereas these are independent of  the  linear rise  (Fig. \ref{para1}).
Neither WD CCM models, neither the sample of  \citet{Vaetal04}  show any correlation with the bump 
position,  moreover their bump height and  linear rise are correlated, pointing out that the
mechanisms  working in the environments of anomalous curves are different from those in the environments of  normal curves.
Type C  extinction  curves require, indeed,   dust abundances lower than normal lines, especially of  silicate grains. Moreover,  a number of small grains of both  silicate and carbonaceous dust
larger than expected for normal curves  with the same R$_V$ value are derived.
Such properties are different from those expected for A and B types too, not only in term of abundances, but also in term of small-to-large grain size ratios. B type curves, in particular, correspond to
sight lines  with the lowest small-to-large grain size ratio of silicate dust,  also compared  with  WD CCM curves.
Carbonaceous grains  do not present a clear difference in such a ratio between anomalous types. However, their ratios  are almost six times larger than those of WD  normal lines with the same R$_V$.

The anomalous extinction properties here analyzed should arise along sight lines where shocks and high velocity flows perturb the physical state of the interstellar medium living their signature on the dust properties. Evidences in this sense have been reported in the previous section.
Shocks and grain-grain collisions should modify the  size distribution of the dust, increasing the number of small grains or, for relatively high velocity shocks, destroy them, reducing the amount of the dust trapped into the grains \citep{{Guetal09},{Jones091},{Jones092},{Dr09}}.

As discussed in \citetalias{MB08}, in order to interpret the results derived for B anomalous lines of sight, both 
a lower small-to-large  silicate grain size ratio and a larger ratio for carbonaceous ones, compared with 
the normal curves, are required.  This can be obtained with a relatively high velocity  shock implying a
sputtering or vaporisation process which sensibly destroys small silicate grains while it 
produces only a partial destruction of  carbonaceous ones with the consequence of increasing the number of  
smaller particles of such a component.
Along A and C  sight lines  slower  velocity shocks than along B sight lines should be required
in order to produce only a partial destruction of large size grains of both the
dust components  increasing the number of smaller particles. 
Moreover,  type  C extinction properties point towards environments where the abundance of dust trapped into the grains is  about two times less than that characterizing  type A and B  environments, and three times less than that of normal, CCM, lines.

\begin{acknowledgements}
We thank Anna Geminale who help us to extract our  sample and an anonymous referee whose useful suggestions help us to improve the paper.
\end{acknowledgements}

\bibliographystyle{aa}

\addtocounter{table}{-9}
\clearpage\onecolumn

\longtab{1}{
\begin{longtable}{llrrrl}
\label{anom0}\\
\caption{Properties of type C anomalous curves }\\
\hline\hline
Name   &Sp.& E(B-V) & V & R$_V$\\
\hline
\endfirsthead
\caption{continued.}\\
\hline\hline
Name   &Sp.& E(B-V) & V & R$_V$\\
\hline
\endhead
\hline
\endfoot
HD\,1337      &O9III     &0.34& 5.90&0.60$\pm$0.18\\
HD\,14357     &B2II       &0.56& 8.53&2.31$\pm$0.21\\
HD\,14707     &B0.5III   &0.83& 9.89 &4.00$\pm$0.23\\
HD\,14734     &B0.5V      &0.55& 9.34&2.20$\pm$0.33\\
HD\,37061     &B1V        &0.52& 6.83&4.50$\pm$0.38\\
HD\,37767     &B3V         &0.35& 8.94&2.87$\pm$0.39\\
HD\,46867     &B0.5III/IV&0.50& 8.30&2.59$\pm$0.26\\
HD\,137569    &B5III     &0.40& 7.86&1.10$\pm$0.18\\ 
HD\,156233    &O9.5II    &0.72& 9.08&2.92$\pm$0.19\\
HD\,164492    &O7/8III    &0.31& 7.63&4.20$\pm$0.62\\
HD\,191396    &B0.5II    &0.53& 8.13&2.65$\pm$0.24\\
HD\,191611    &B0.5III   &0.65& 8.59&2.81$\pm$0.20\\ 
HD\,282622    &B1/2V     &0.56& 9.66&5.41$\pm$0.43\\ 
HD\,344784    &B0IV     &0.86& 9.34&3.01$\pm$0.16\\ 
HD\,392525    &B1/2IV/V  &0.50&10.35&4.54$\pm$0.54\\
BD$+$23\,3762   &B0.5III   &1.05& 9.29&2.47$\pm$0.12\\
BD$+$52\,3122   &B2II      &0.56& 9.31&5.35$\pm$0.59\\
BD$+$55\,2770   &B1/2III   &0.60& 9.70&2.90$\pm$0.19\\ 
BD$+$56\,586  &B1V    &0.51 & 9.94 & 2.59$\pm$0.39\\ 
BD$+$57\,252    &B3V       &0.52& 9.50&2.97$\pm$0.27\\
BD$+$59\,273    &B2III     &0.46& 9.08&2.65$\pm$0.28\\
BD$+$63\,89     &B1Ib      &0.79& 9.50&2.95$\pm$0.17\\ 
HD\,2619      &B0.5III    &0.85& 8.31&2.55$\pm$0.15\\
HD\,21455     &B7V         &0.26& 6.24&3.17$\pm$0.59\\
HD\,28446    &B0III       &0.46& 5.78&2.46$\pm$0.26\\
HD\,38658     &B3II        &0.40& 8.35&2.62$\pm$0.32\\
HD\,41831     &B3V         &0.36& 9.16&2.86$\pm$0.38\\
HD\,54439     &B2III       &0.28& 7.70&2.13$\pm$0.41\\
HD\,73420    &B2II/III    &0.37& 8.86&2.47$\pm$0.32\\
HD\,78785    &B2II        &0.76& 8.61&2.55$\pm$0.17\\
HD\,96042     &O9.5V       &0.48& 8.23&1.97$\pm$0.24\\
HD\,141318    &B2II       &0.30& 5.73&1.95$\pm$0.18\\
HD\,149452    &O9V       &0.90& 9.05&3.20$\pm$0.16\\
HD\,152245    &B0III      &0.42& 8.37&2.25$\pm$0.29\\
HD\,152853    &B2II/III   &0.37& 7.94&2.50$\pm$0.33\\
HD\,161061    &B2III      &1.01& 8.47&2.92$\pm$0.14\\
HD\,168021    &B0Ib       &0.55& 6.84&3.15$\pm$0.27\\
HD\,168137    &O8V        &0.73& 8.85&2.97$\pm$0.22\\
HD\,168785    &B2III     &0.30& 8.48&2.03$\pm$0.37\\
HD\,168894    &B1I       &0.90& 9.38&2.92$\pm$0.16\\
HD\,173251    &B1II      &0.93& 9.09&2.64$\pm$0.14\\
HD\,194092    &B0.5III    &0.41& 8.26&2.50$\pm$0.30\\
HD\,211880    &B0.5V      &0.60& 7.75&2.65$\pm$0.22\\
HD\,216248    &B3II       &0.64& 9.89&2.85$\pm$0.22\\
HD\,217035   &B0V        &0.76& 7.74&2.77$\pm$0.18\\
HD\,218323   &B0III      &0.90& 7.63&2.55$\pm$0.15\\
HD\,226868    &B0Ib       &1.08& 8.89&3.20$\pm$0.14\\
HD\,229049    &B2III      &0.72& 9.62&2.65$\pm$0.18\\
HD\,248893    &B0II/III  &0.74& 9.69&2.81$\pm$0.18\\
HD\,252325    &B2V        &0.70&10.79&3.63$\pm$0.20\\
HD\,253327    &B0.5V      &0.86&10.76&3.09$\pm$0.17\\
HD\,326327    &B1V        &0.53& 9.75&3.07$\pm$0.22\\
HD\,344894    &B2III      &0.57& 9.61&2.50$\pm$0.22\\
HD\,345214    &B5III     &0.39& 9.34&2.45$\pm$0.31\\ 
BD$+$45\,3341&B1II    &0.74& 8.73&2.46$\pm$0.17\\  
BD$+$52\,3135   &B3II      &0.53& 9.62&2.97$\pm$0.27\\
BD$+$58\,310    &B1V       &0.51&10.17&5.65$\pm$0.47\\
BD$+$59\,2829   &B2II      &0.70& 9.84&3.96$\pm$0.26\\
BD$+$60\,2380   &B2III     &0.63& 9.04&2.77$\pm$0.22\\
BD$+$62\,338 &B3II    &0.41& 9.22&2.55$\pm$0.31\\
BD$+$62\,2142&B3V     &0.60& 9.04&2.81$\pm$0.22\\
BD$+$62\,2154&B1V     &0.77& 9.33&2.75$\pm$0.17\\
BD$+$62\,2353&B3II  &0.44& 9.87&2.31$\pm$0.26\\
BD$+$63\,1964&B0II  &1.01& 8.46&2.70$\pm$0.20\\
\hline
\end{longtable}
}

\longtab{2}{
\begin{longtable}{lllllllllll}
\label{parmod}\\
\caption{Best-fit parameters of WD grain size distributions of type C curves}\\
\hline\hline
Name&$10^5b_C $ & $\alpha_g$&$\beta_g$&$a_{t,g}$&$a_{c,g}$&$C_g$&$\alpha_s$&$\beta_s$&$a_{t,s}$ &$C_s$\\
     &      &       &     &$(\mu m)$&$(\mu m)$&  &  &           &$(\mu m)$&\\
\hline
\endfirsthead
\caption{continued}\\
  \hline\hline
Name&$10^5b_C $ & $\alpha_g$&$\beta_g$&$a_{t,g}$&$a_{c,g}$&$C_g$&$\alpha_s$&$\beta_s$&$a_{t,s}$ &$C_s$\\
     &      &       &     &$(\mu m)$&$(\mu m)$&  &  &           &$(\mu m)$&\\
\hline
\endhead
\hline
\endfoot
HD\,14357&0.80&11.4&9.39&0.053&0.020&2.50$\times 10^{-15}$&-2.15&-0.05&0.050&2.98$\times 10^{-13}$ \\
HD\,14707&1.00&-1.50&0.30&0.004&0.105&5.00$\times 10^{-12}$ &-1.20&-2.00&0.055&1.73$\times 10^{-13}$ \\
HD\,14734&1.00&-2.79&0.17&0.007&0.560&2.50$\times 10^{-14}$&-2.30&-0.75&0.16&1.03$\times 10^{-13}$ \\
HD\,37061&4.00&13.7&0.98&0.052&0.024&5.60$\times 10^{-13}$&-2.20&3.40&0.053&3.10$\times 10^{-13}$ \\
HD\,37767&3.00&9.70&-1.20&0.043&0.029&7.50$\times 10^{-14}$ &-2.10&2.40&0.045&3.80$\times 10^{-13}$ \\
HD\,46867&3.50&11.0&4.13&0.048&0.023&1.75$\times 10^{-14}$&-1.38&1.52&0.049&9.96$\times 10^{-13}$\\
HD\,156233&4.80&11.5&3.00&0.053&0.023&1.85$\times 10^{-14}$ &-1.50&1.00&0.048&9.60$\times 10^{-13}$\\
HD\,164492&2.70&7.50&-0.04&0.050&0.030&2.80$\times 10^{-13}$ &-1.30&4.00&0.045&3.76$\times 10^{-13}$ \\
HD\,191396&0.65&12.4&10.9&0.054&0.020&2.00$\times10^{-15}$&-1.50&-0.21&0.050&2.98$\times 10^{-13}$\\
HD\,191611&1.00&11.4&5.91&0.053&0.022&3.30$\times 10^{-14}$&-2.35&-0.05&0.053&2.96$\times 10^{-13}$\\
HD\,282622&0.86&-1.65&0.53&0.004&0.125&3.50$\times 10^{-12}$&-1.30&-3.20&0.059&1.53$\times 10^{-12}$\\
HD\,344784&4.25&11.0&0.50&0.051&0.024&5.60$\times 10^{-14}$&-2.20&2.70&0.053&3.10$\times 10^{-13}$\\ 
HD\,392525&1.80&11.0&5.13&0.062&0.022&1.75$\times10^{-14}$&-1.38&-0.10&0.061&6.96$\times 10^{-13}$\\
BD$+$23\,3762&4.50&7.00&1.45&0.055&0.024&1.03$\times 10^{-13}$&-2.15&0.15&0.070&3.76$\times 10^{-13}$\\
BD$+$52\,3122&0.10&-1.73&0.30&0.041&0.012&5.00$\times 10^{-12}$&-1.30&-4.00&0.055&1.53$\times 10^{-12}$\\
BD$+$55\,2770&0.80&13.4&11.4&0.057&0.019&1.70$\times 10^{-15}$&-1.85&-0.15&0.050&2.98$\times 10^{-13}$\\
BD$+$57\,252&0.65&13.4&11.9&0.058&0.019&1.80$\times 10^{-15}$&-1.50&-0.21&0.050&2.98$\times 10^{-13}$\\
BD$+$59\,273&4.80&11.5&3.0&0.051&0.023&1.85$\times 10^{-14}$&-1.80&1.80&0.048&9.60$\times 10^{-13}$\\
BD$+$63\,89&4.50&9.70&1.50&0.052&0.024&5.45$\times 10^{-14}$&-2.20&3.40&0.052&3.10$\times 10^{-13}$\\
HD\,2619&5.00&8.70&-0.50&0.052&0.025&2.70$\times 10^{-13}$ &-2.60&0.55&0.069&2.88$\times 10^{-13}$ \\
HD\,21455&3.50&8.70&2.52&0.054&0.024&8.30$\times 10^{-14}$ &-2.90&-0.14&0.069&4.82$\times 10^{-13}$ \\
HD\,28446&3.80&9.00&-0.95&0.044&0.029&1.10$\times 10^{-13}$ &-2.40&2.20&0.050&3.10$\times 10^{-13}$ \\
HD\,38658&3.00&9.70&-1.30&0.041&0.029&6.50$\times 10^{-14}$ &-1.90&2.80&0.045&3.80$\times 10^{-13}$ \\
HD\,41831&1.00&11.7&5.91&0.053&0.023&3.30$\times 10^{-14}$ &-0.75&-0.10&0.055&4.89$\times 10^{-13}$ \\
HD\,54439&3.80&7.90&-0.55&0.046&0.026&2.70$\times 10^{-13}$ &-2.00&0.90&0.069&2.40$\times 10^{-13}$ \\
HD\,73420&3.20&7.90&-0.55&0.049&0.026&2.70$\times 10^{-13}$ &-2.40&0.32&0.069&1.60$\times 10^{-13}$ \\
HD\,78785&3.20&7.90&-0.65&0.052&0.026&2.70$\times 10^{-13}$ &-2.60&0.72&0.069&2.70$\times 10^{-13}$ \\
HD\,96042&4.20&7.90&-0.65&0.045&0.026&2.70$\times 10^{-13}$ &-2.80&0.93&0.069&2.40$\times 10^{-13}$ \\
HD\,141318&4.20&7.90&-0.65&0.043&0.026&2.70$\times 10^{-13}$ &-2.60&0.32&0.069&2.40$\times 10^{-13}$ \\
HD\,149452&3.50&7.25&.024&0.054&0.02&2.60$\times 10^{-13}$&-2.55&0.50&0.067&2.88$\times 10^{-13}$\\
HD\,152245&4.20&7.95&-0.65&0.048&0.026&2.70$\times 10^{-13}$ &-2.60&0.28&0.069&2.40$\times 10^{-13}$ \\
HD\,152853&3.70&7.90&-0.65&0.050&0.026&2.70$\times 10^{-13}$ &-2.00&0.32&0.069&2.40$\times 10^{-13}$ \\
HD\,161061&4.80&11.5&3.00&0.053&0.023&1.75$\times 10^{-14}$ &-1.50&1.00&0.049&9.60$\times 10^{-13}$ \\
HD\,168021&0.32&4.98&1.86&0.078&0.106&5.70$\times 10^{-14}$ & 0.35&-1.60&0.030&6.76$\times 10^{-13}$ \\
HD\,168137&1.10&11.4&5.91&0.053&0.023&3.30$\times 10^{-15}$ &-2.05&-0.14&0.053&2.98$\times 10^{-13}$ \\
HD\,168785&1.00&11.0&5.91&0.041&0.025&3.10$\times10^{-14}$&-2.00&-0.11&0.073&3.30$\times 10^{-13}$ \\
HD\,168894&4.80&8.80&-0.50&0.051&0.027&2.70$\times 10^{-13}$ &-2.70&0.85&0.069&3.10$\times 10^{-13}$ \\
HD\,173251&0.40&0.70&1.52&0.080&0.012& 5.70$\times10^{-14}$&-0.15&-0.26&0.042&5.38$\times 10^{-13}$\\
HD\,194092&4.20&7.90&-0.60&0.050&0.026&2.70$\times 10^{-13}$ &-2.60&0.32&0.069&2.40$\times 10^{-13}$ \\
HD\,211880&3.20&8.30&-0.55&0.051&0.026&2.37$\times 10^{-13}$ &-2.30&1.40&0.062&2.50$\times 10^{-13}$ \\
HD\,216248&3.30&5.18&3.25&0.066&0.019&1.70$\times 10^{-13}$ &-2.50&-0.20&0.057&6.40$\times 10^{-13}$ \\
HD\,217035&3.30&7.90&-0.65&0.053&0.026&2.70$\times 10^{-13}$ &-1.80&0.83&0.069&2.30$\times 10^{-13}$ \\
HD\,218323&3.30&7.90&-0.65&0.052&0.026&2.70$\times 10^{-13}$ &-2.50&0.83&0.069&2.30$\times 10^{-13}$ \\
HD\,226868&4.15&8.60&-0.50&0.053&0.027&2.70$\times 10^{-13}$ &-2.60&0.86&0.069&2.88$\times 10^{-13}$ \\
HD\,229049&4.00&7.90&-0.65&0.053&0.026&2.70$\times 10^{-13}$ &-2.60&0.32&0.069&2.00$\times 10^{-13}$ \\
HD\,248893&2.90&11.4&5.92&0.048&0.025&3.10$\times 10^{-14}$&-2.50&-0.11&0.071&3.30$\times 10^{-13}$\\
HD\,252325&2.48&11.0&5.13&0.058&0.022&1.75$\times 10^{-14}$ &-1.38&-0.80&0.059&6.69$\times 10^{-13}$ \\
HD\,253327&1.10&11.4&5.91&0.054&0.023&3.30$\times 10^{-15}$ &-0.40&-0.41&0.056&5.10$\times 10^{-13}$ \\
HD\,326327&3.20&8.49&-0.65&0.053&0.026&2.70$\times 10^{-13}$ &-2.90&0.32&0.070&2.40$\times 10^{-13}$ \\
HD\,344894&5.50&8.70&-0.50&0.051&0.025&2.70$\times 10^{-13}$ &-2.60&0.32&0.070&2.88$\times 10^{-13}$ \\
HD\,345214&0.90&11.9&9.39&0.053&0.020&2.50$\times10^{-15}$&-1.20&-0.21&0.050&2.98$\times 10^{-13}$\\
BD$+$45\,3341&3.20&7.90&-0.65&0.051&0.026&2.70$\times 10^{-13}$ &-2.60&1.85&0.069&2.00$\times 10^{-13}$ \\
BD$+$52\,3135&4.00&8.70&-0.50&0.051&0.026&2.70$\times 10^{-13}$&-2.80&0.32&0.069&2.88$\times 10^{-13}$\\
BD$+$58\,310&0.60&-1.65&0.53&0.408&0.124&3.50$\times 10^{-12}$&-1.30&-3.50&0.058&1.53$\times 10^{-12}$\\
BD$+$59\,2829&0.15&-1.70&0.08&0.004&0.111&6.50$\times 10^{-12}$&-1.30&-6.00&0.055&1.10$\times 10^{-12}$\\
BD$+$60\,2380&4.50&8.70&-0.50&0.051&0.026&2.70$\times 10^{-13}$&-2.70&0.32&0.069&2.88$\times 10^{-13}$\\
BD$+$62\,338&3.00&9.70&-1.75&0.034&0.031&4.70$\times 10^{-14}$ &-1.80&1.50&0.050&3.80$\times 10^{-13}$ \\
BD$+$62\,2142&3.80&9.70&-1.75&0.039&0.032&5.00$\times 10^{-14}$ &-1.20&1.00&0.050&6.80$\times 10^{-13}$ \\
BD$+$62\,2154&3.00&9.70&-1.75&0.042&0.031&5.00$\times 10^{-14}$ &-1.60&1.00&0.057&3.80$\times 10^{-13}$ \\
BD$+$62\,2353&3.00&9.70&-1.30&0.039&0.029&5.00$\times 10^{-14}$ &-1.90&2.80&0.045&3.80$\times 10^{-13}$ \\
BD$+$63\,1964&1.15&10.0&4.075&0.052&0.030&6.33$\times 10^{-15}$ &-0.75&-0.20&0.055&4.98$\times 10^{-13}$ \\
\end{longtable}
}

\longtab{4}{
\begin{longtable}{llllllll}
\label{resanom1}\\
\caption{Properties of the dust trapped into the grains along type C curves}\\
\hline\hline
Name&$\frac{\rho_d}{\rho_H}$&$\frac{C}{C_{\odot}}$&$\frac{Si}{Si_{\odot}}$&$\frac{R_C}{10^{4}}$&
$\frac{R_{Si}}{10^{2}}$&$R_V$&$\frac {E(B-V)}{N_H }\times10^{-22}$\\
     &      &       &   &   &   &    &(mag\,cm$^{2}$)  \\
\hline
\endfirsthead
\caption{continued.}\\
\hline\hline
Name&$\frac{\rho_d}{\rho_H}$&$\frac{C}{C_{\odot}}$&$\frac{Si}{Si_{\odot}}$&$\frac{R_C}{10^{4}}$&
$\frac{R_{Si}}{10^{2}}$&$R_V$&$\frac {E(B-V)}{N_H }\times10^{-22}$\\
     &      &       &   &   &   &    & (mag\,cm$^{2}$)  \\
\hline
\endhead
\hline
\endfoot
HD\,14357    &0.07&0.13& 0.06&18   & 1.4 & 2.12& 0.09 \\
HD\,14707    &0.17&0.36& 0.12& 3.6 & 0.25& 3.60& 0.42 \\
HD\,14734   & 0.36 &0.21&0.49&19.4  & 0.46& 1.98& 0.62 \\
HD\,37061    &0.92&2.81& 0.23& 8.8 & 0.8 & 3.82& 2.57 \\
HD\,37767    &0.24&0.54&0.14 &17.6 &0.73 & 2.53& 0.91 \\
HD\,46867    &0.41&0.87& 0.26&14.5 & 0.2 & 2.31& 1.71 \\
HD\,156233   &0.34&0.77& 0.20&19   & 0.2 & 2.58& 1.28 \\
HD\,164492   &0.44&1.28& 0.15&11   & 0.09& 3.68& 1.75 \\
HD\,191396   &0.06&0.15& 0.04&16   & 0.3 & 2.37& 0.28 \\
HD\,191611   &0.10&0.18& 0.09&18   & 2.2 & 2.46& 0.32 \\
HD\,282622   &0.17&0.39& 0.10& 3.8 & 0.35 & 4.93& 0.33 \\
HD\,344784   &0.36&0.85& 0.20&17   & 0.9 & 2.66& 1.02 \\
HD\,392525   &0.23&0.48& 0.15&16.5   & 0.2 & 3.86& 0.62 \\
BD$+$23\,3762&0.31&0.60& 0.23&20   & 1.3 & 2.15& 1.07 \\
BD$+$52\,3122&0.10&0.23& 0.07& 0.9 & 0.4 & 4.82& 0.20 \\
BD$+$55\,2770&0.07&0.14& 0.05&18   & 0.74& 2.56& 0.22 \\
BD$+$57\,252 &0.06&0.13& 0.04&17   & 0.3 & 2.64& 0.22 \\
BD$+$59\,273 &0.42&0.83& 0.30&17.5 & 0.4 & 2.42& 1.49 \\
BD$+$63\,89  &0.37&0.83& 0.22&18   & 0.8 & 2.54& 1.48 \\
HD\,2619     &0.40&0.72& 0.32&19   & 3.5 & 2.34& 1.28 \\
HD\,21455    &0.74&1.07&0.71 &14   & 7.9 & 2.81& 1.84 \\
HD\,28446    &0.26&0.55&0.18 &18.7 &1.6  & 2.20& 1.01 \\
HD\,38658    &0.23&0.51&0.14 &17.6 &0.43 & 2.31& 0.92 \\
HD\,41831    &0.09&0.20&0.06 &17.3 &.06  & 2.56& 1.14 \\
HD\,54439    &0.27&0.59&0.17 &17.5 &0.82 & 1.98& 1.21 \\
HD\,73420    &0.23&0.54&0.13 &17.5 &2.3  & 2.24& 1.07 \\
HD\,78785    &0.34&0.50&0.32 &18.5 &3.4  & 2.29& 0.92 \\
HD\,96042    &0.41&0.60&0.40 &18   & 5.2 & 1.87& 2.21 \\
HD\,141318   &0.33&0.63& 0.25&16   & 3.6 & 1.77& 1.36 \\
HD\,149452   &0.37&0.70& 0.28&18   & 3.1 & 2.76& 1.14 \\
HD\,152245   &0.31&0.57&0.25 &19   & 3.7 & 2.02& 1.11 \\
HD\,152853   &0.23&0.52&0.13 &19   & 0.91& 2.19& 0.96 \\
HD\,161061   &0.45&0.82& 0.35&19   & 3.3 & 2.58& 1.37 \\
HD\,168021   &0.02&0.06& 0.01&18   & 0.02& 2.74& 0.09 \\ 
HD\,168137   &0.09&0.20& 0.06&18   & 1.15& 2.52& 0.35 \\
HD\,168785   &0.33&0.82& 0.16&18   & 1.0 & 1.91& 1.69 \\
HD\,168894   &0.51&0.85& 0.43&18   & 4.2 & 2.56& 1.46 \\ 
HD\,173251   &0.04&0.07& 0.03&14   & .02 & 2.35& 0.13 \\
HD\,194092   &0.28&0.54&0.25 &21   & 3.6 & 2.18& 0.98 \\
HD\,211880   &0.29&0.56&0.20 &18   & 1.50& 2.40& 1.09 \\
HD\,216248   &0.32&0.55&0.27 &18.2 & 3.2 & 2.52& 0.94 \\
HD\,217035   &0.22&0.49&0.14 &19   & 0.5 & 2.44& 0.85 \\
HD\,218323   &0.29&0.50&0.25 &19   & 2.6 & 2.30& 0.91 \\
HD\,226868   &0.43&0.74& 0.35&18   & 3.3 & 2.78& 1.13 \\
HD\,229049   &0.27&0.52&0.21 &20   & 3.6 & 2.34& 0.89 \\
HD\,248893   &0.32&0.51& 0.28&18   & 3.1 & 2.47& 0.92 \\
HD\,252325   &0.21&0.53& 0.10&17.4 & 0.30& 3.15& 0.78 \\
HD\,253327   &0.08&0.19& 0.05&18.5 & 0.03& 2.67& 0.32 \\
HD\,326327   &0.31&0.57&0.25 &18   & 3.6 & 2.61& 0.94 \\
HD\,344894   &0.40&0.76& 0.30&19   & 3.6 & 2.21& 1.38 \\
HD\,345214   &0.06&0.15& 0.03&18   & 0.2 & 2.06& 0.29 \\
BD$+$45\,3341&0.33&0.50&0.32 &18.3 &2.84 & 2.22& 0.96 \\
BD$+$52\,3135&0.47&0.78& 0.40&17   & 5.8 & 2.64& 1.34 \\
BD$+$58\,310 &0.16&0.37& 0.09& 2.8 & 0.4 & 5.04& 0.31 \\
BD$+$59\,2829&0.05&0.11& 0.04& 1.2 & 0.5 & 3.55& 0.12 \\
BD$+$60\,2380&0.43&0.76& 0.35&18   & 4.6 & 2.45& 1.32 \\
BD$+$62\,338 &0.21&0.47&0.12 &18.3 &0.44 & 2.33& 0.84 \\
BD$+$62\,2142&0.26&0.62&0.15 &18.4 &0.12 & 2.50& 1.03 \\
BD$+$62\,2154&0.21&0.45&0.13 &19.1 &0.31 & 2.43& 0.76\\
BD$+$62\,2353&0.22&0.49&0.14 &17.4 &0.43 & 2.09& 0.94 \\
BD$+$63\,1964&0.24&0.46&0.19 &18.4 &0.46 & 2.42& 0.81 \\
\end{longtable}
}

\longtab{6}{
\begin{longtable}{lllllll}
\label{FMpar}\\
\caption{FM parameters of  type C anomalous curves}\\
\hline\hline
Name& $c_1$ & $c_2$ & $c_3$&$c_4$&$x_0$ & $\gamma$\\
\hline
\endfirsthead
\caption{continued.}\\
\hline\hline
Name& $c_1$ & $c_2$ & $c_3$&$c_4$&$x_0$ & $\gamma$\\
\hline
\endhead
\hline
\endfoot
HD\,14357      & -1.492 & 0.952 &  1.593 &0.188& 4.562& 0.801 \\
HD\,14707      & -1.231 & 1.076 &  2.736 &0.126& 4.524& 0.947 \\
HD\,14734      & -2.028 & 1.633 &  2.731 &0.288& 4.547& 0.921 \\
HD\,37061      &  0.547 & 0.102 &  1.150 &0.110& 4.469& 0.927 \\
HD\,37767      & -0.482 & 0.636 &  2.212 &0.174& 4.574& 0.886\\
HD\,46867      & -0.204 & 0.508 &  1.998 &0.138& 4.571& 0.915 \\
HD\,156233     & -0.679 & 0.708 &  3.000 &0.185& 4.590& 0.959 \\
HD\,164492     &  0.773 & 0.092 &  1.700 &0.206& 4.512& 0.907\\
HD\,191396     & -0.238 & 0.505 &  1.726 &0.114& 4.565& 0.905  \\
HD\,191611     & -1.688 & 0.989 &  1.508 &0.235& 4.552& 0.779  \\
HD\,282622     & -1.531 & 1.147 &  2.612 &0.136& 4.503& 0.927 \\
HD\,344784     & -0.187 & 0.510 &  2.419 &0.171& 4.575& 0.940 \\
HD\,392525     & -1.1830 & 0.812 &  3.867 &0.008& 4.615&1.133 \\ 
BD$+$23\,3762  & -0.502 & 0.731 &  3.698 &0.262& 4.593& 0.980\\
BD$+$52\,3122  & -2.258 & 1.308 &  1.704 &0.063& 4.411& 0.916\\
BD$+$55\,2770  & -0.700 & 0.719 &  1.597 &0.133& 4.550& 0.778 \\
BD$+$57\,252   & -0.688 & 0.661 &  2.195 &0.132& 4.582& 0.920 \\
BD$+$59\,273   & -0.807 & 0.744 &  2.786 &0.146& 4.593& 0.968\\
BD$+$63\,89    & -0.397 & 0.595 &  2.577 &0.164& 4.582& 0.949\\
HD\,2619       & -0.846 & 0.811 &  2.744 &0.312& 4.573& 0.894\\
HD\,21455      & -2.279 & 1.050 &  0.662 &0.367& 4.452& 0.679\\
HD\,28446      & -0.534 & 0.693 &  2.581 &0.231& 4.576& 0.893\\
HD\,38658      & -0.500 & 0.657 &  2.343 &0.147& 4.584& 0.913\\
HD\,41831      & -0.771 & 0.688 &  4.039 &0.068& 4.625& 1.173\\
HD\,54439      &  0.269 & 0.461 &  3.175 &0.190& 4.586& 1.014\\
HD\,73420      &  0.306 & 0.410 &  2.529 &0.254& 4.563& 0.920\\
HD\,78785      & -1.747 & 1.109 &  2.286 &0.258& 4.575& 0.878\\
HD\,96042      & -1.318 & 0.979 &  1.909 &0.311& 4.557& 0.827\\
HD\,141318     & -0.312 & 0.652 &  2.310 &0.289& 4.561& 0.875\\
HD\,149452     & -0.845 & 0.757 &  2.127 &0.248& 4.561& 0.880 \\
HD\,152245     &  0.524 & 0.722 &  2.545 &0.311& 4.567& 0.881\\
HD\,152853     &  0.232 &0.477  &  3.594 &0.253& 4.587& 0.991\\
HD\,161061     & -0.898 & 0.812 &  2.650 &0.281& 4.573& 0.904\\
HD\,168021     &  0.395 & 0.332 &  2.134 &0.234& 4.547& 0.843\\
HD\,168137     & -0.654 & 0.664 &  2.016 &0.202& 4.564& 0.865\\
HD\,168785     &  0.646 & 0.316 &  2.713 &0.230& 4.569& 0.951 \\
HD\,168894     & -1.274 & 0.915 &  2.105 &0.283& 4.563& 0.859\\
HD\,173251     & -1.067 & 0.913 &  2.734 &0.037& 4.605& 0.999 \\
HD\,194092     & -0.435 & 0.698 &  3.035 &0.353& 4.572& 0.896\\
HD\,211880     & -0.364 & 0.612 &  2.587 &0.174& 4.579& 0.967\\
HD\,216248     & -0.707 & 0.634 &  6.253 &0.067& 4.660& 1.374\\
HD\,217035     &  2.166 & 0.544 &  4.369 &0.221& 4.603& 1.064\\
HD\,218323     & -1.061 & 0.892 &  2.809 &0.246& 4.583& 0.931\\
HD\,226868     & -1.266 & 0.921 &  2.456 &0.265& 4.572& 0.895\\
HD\,229049     & -0.625 & 0.759 &  3.136 &0.366& 4.574& 0.895\\
HD\,248893     & -1.400 & 0.941 &  1.870 &0.268& 4.559& 0.835\\
HD\,252325     &  0.062 & 0.421 &  2.895 &0.198& 4.576& 0.971\\
HD\,253327     & -0.486 & 0.670 &  2.797 &0.266& 4.574& 0.923\\
HD\,326327     & -1.001 & 0.818 &  2.218 &0.294& 4.561& 0.863\\
HD\,344894     & -0.503 & 0.702 &  2.780 &0.331& 4.570& 0.890 \\
HD\,345214     &  0.083 & 0.445 &  2.572 &0.167& 4.581& 0.956 \\
BD$+$45\,3341     & -1.578 & 1.068 &  2.698 &0.199& 4.591& 0.951\\
BD$+$52\,3135  & -1.306 & 0.878 &  1.572 &0.333& 4.537& 0.792\\
BD$+$58\,310   & -1.360 & 1.069 &  2.533 &0.147& 4.484& 0.936 \\
BD$+$59\,2829  & -1.660 & 1.196 &  2.209 &0.166& 4.464& 0.912 \\
BD$+$60\,2380  & -0.921 & 0.794 &  2.073 &0.320& 4.555& 0.842\\
BD$+$62\,2353  & -0.427 & 0.648 &  2.269 &0.142& 4.583& 0.910\\
BD$+$62\,338   & -0.301 & 0.608 &  2.966 &0.178& 4.589& 0.961\\
BD$+$62\,2142  & -0.254 & 0.586 &  3.578 &0.164& 4.602& 1.023\\
BD$+$62\,22154 & -0.466 & 0.671 &  4.130 &0.178& 4.605& 1.053\\
BD$+$63\,1964  & -0.950 & 0.826 &  4.635 &0.121& 4.620& 1.134\\
\end{longtable}
}

\end{document}